%% file: main.tex
\documentclass[preprint, 12pt]{elsarticle}

\usepackage[utf8]{inputenc}
\usepackage[english]{babel}
\usepackage{amssymb, amsfonts, amsbsy, amsmath, amsthm}
\usepackage{mathrsfs}
\usepackage[T1]{fontenc} 
\usepackage{graphicx}
\usepackage{geometry}
\usepackage{comment}
\usepackage{bbm}
\usepackage{float}
\usepackage{url}
\usepackage{setspace}
\usepackage{color}
\usepackage{subcaption}
\usepackage{nomencl}
\usepackage[font=small,labelfont=bf,tableposition=top]{caption}
\usepackage[colorlinks=true, allcolors=blue]{hyperref}
\usepackage{orcidlink}

\DeclareCaptionLabelFormat{andtable}{#1~#2  \&  \tablename~\thetable}

\newdefinition{rmk}{Remark}
\newproof{pf}{Proof}
\newproof{pot}{Proof of Theorem \ref{thm2}}

\renewcommand{\P}{{\mathbb P}}
\newcommand{\E}{{\mathbb E}}
\newcommand{\N}{{\mathbb N}}

\title{\textit{e}-Values for Real-Time Residential Electricity Demand Forecast Model Selection}

\author[inst1]{Fabian Backhaus
\orcidlink{0009-0009-8995-4444}\corref{cor1}}

\author[inst1]{Karoline Brucke\orcidlink{0000-0002-4510-8969}}

\author[inst2]{Peter Ruckdeschel\orcidlink{0000-0001-7815-4809}}

\author[inst1]{Sunke Schlüters\orcidlink{0000-0002-2186-812X}}

\cortext[cor1]{Corresponding author \textit{Email adress:} \texttt{fabian.backhaus@dlr.de} (Fabian Backhaus)}

\affiliation[inst1]{
    organization={DLR-Institute of Networked Energy Systems},
    addressline={Carl-von-Ossietzky-Str. 15}, 
    city={Oldenburg},
    postcode={26129}, 
    country={Germany}
    }
\affiliation[inst2]{
    organization={Carl-von-Ossietzky University Oldenburg},
    addressline={Ammerländer Heerstraße 114-118}, 
    city={Oldenburg},
    postcode={26129}, 
    country={Germany}
    }
\journal{Energy and Buildings}

\begin{document}
\begin{highlights}
\item ntroduction of $e$-values for sequential testing in the energy demand forecasting domain.
\item Combines forecasts based on $e$-values to generate predictions with guarantees.
\item Enhances performance with $e$-value model selection on electricity demand forecasts.
\item Sequentially compares continuous forecasts, controlling for familywise error rates.
\item Extends $e$-selection approach to unbounded scores, such as Mean Absolute Error.

\end{highlights}

\begin{abstract}
With the growing number of forecasting techniques and the increasing significance of forecast-based operation - particularly in the rapidly evolving energy sector - selecting the most effective forecasting model has become a critical task.
Given the dynamic nature of energy forecasting, it is highly advantageous to 
assess the superiority of forecasting models not only retrospectively but continuously in real-time as new data and evidence becomes available, while simultaneously providing strong probabilistic guarantees for these decisions. 
In this work, we show that this can be achieved through the mathematical concept of $e$-values, which has recently gained massive attention in the field of statistics. 
It allows for unified construction principles for powerful tests and accurate statistical decisions, which can be evaluated at any chosen time points while maintaining an overall probabilistic error control. We extend the use of $e$-values by developing a simple persistence approach that dynamically combines input forecasts to generate new fused predictions. 
To demonstrate the performance of our method we apply it to electricity demand forecasts based on different artificial intelligence based models.  
Our results indicate that $e$-values are able to improve the accuracy and reliability of forecasts in a dynamic environment, offering a valuable tool for real-time decision-making in the energy sector. 
\end{abstract}


\maketitle
\makenomenclature
\nomenclature[01]{\(\alpha\)}{Significance level.}
\nomenclature[02]{\(\E_P\)}{Conditional expectation with respect to distribution \(P\).}
\nomenclature[03]{\(\mathcal{F}_t\)}{$\sigma$-Algebra given the information up to time \(t\).}
\nomenclature[04]{\((p_t),~(q_t)\)}{Forecast sequences.}
\nomenclature[05]{\(\mathcal{H}_0(p,q)\)}{Null hypothesis: Forecast \((q_t)\) is not better than \((p_t)\).}
\nomenclature[06]{\(E_\lambda(t),~ E^*_\lambda(t)\)}{$e$-process for \(\mathcal{H}_0(p,q)\) or \(\mathcal{H}_0(q,p)\).}
\nomenclature[07]{\(\mathfrak{p}_t,~ \mathfrak{p}^*_t\)}{$p$-process corresponding to the $e$-process.}
\nomenclature[08]{\(\delta_t,~ \widehat{\delta}_t\)}{(Empirical) score differentials.}
\nomenclature[09]{\(\Delta_t,~ \widehat{\Delta}_t\)}{(Empirical) average score differentials.}
\nomenclature[10]{\(\widetilde{\delta}_t,~ \widetilde{\Delta}_t\)}{Bounded empirical (average) score differentials.}
\nomenclature[11]{\(\omega\)}{Number of past observations. Symbols indexed by \(\omega\) use the last \(\omega\) observations.}
\nomenclature[12]{\(V_{t}\)}{Variance process of the score differentials.}
\nomenclature[13]{\(\psi(\lambda)\)}{$\psi$-function.}
\nomenclature[14]{\(u_{\alpha/2}\)}{Uniform boundary at significance level \(\alpha/2\).}
\nomenclature[15]{\(C_\alpha\)}{$(1-\alpha)$ confidence interval.}
\nomenclature[16]{\(\theta_t\)}{Prediction using the $e$-procedure.}
\nomenclature[17]{\(w_{p,t},~ w_{q,t}\)}{Weights for forecasts \((p_t)\) and \((q_t)\).}
\nomenclature[18]{\(f(x)\)}{Sigmoid function to bound score differentials.}
\printnomenclature

\section{Introduction}
\input{introduction}

\section{Methodology}\label{sec:methodology}
\input{Methodology}

\section{Data Description}\label{data description}
\input{data}

\section{Results}\label{sec:results}
\input{results}

\section{Discussion}\label{sec:discussion}\input{discussion}

\section{Conclusion and Outlook}\label{sec:conclusion}
\input{conclusion}

\bibliographystyle{ieeetr}
\bibliography{literatur}

\section*{CRediT author statement}
\noindent \textbf{Fabian Backhaus}: Writing - Original Draft, Writing -Review \& Editing, Formal analysis, Software, Methodology, Visualization, Validation, Data curation \textbf{Karoline Brucke}: Writing -Original Draft, Supervision, Writing - Review \& Editing, Resources \textbf{Peter Ruckdeschel}: Writing - Original Draft, Writing - Review \& Editing, Methodology, Conceptualization, Supervision  \textbf{Sunke Schlüters}: Supervision, Project administration, Funding acquisition

\section*{Declaration of interests}
\noindent The authors declare that they have no known competing financial interests or personal relationships that could have appeared to influence the work reported in this paper.
\end{document}

%% file: introduction.tex
High quality demand predictions are crucial for energy management in the building context \cite{amasyali2018,deb2017} since prediction-based demand side management is able to harness available decentral flexibility potentials \cite{golmohamadi2022} and enable e.g. load shifting etc. \cite{babatunde2020, gazafroudi2018}. 
Many different forecasting methodologies exist with varying advantages and disadvantages. 
Statistical approaches like standardized load profiles \cite{LPG}, ARIMA \cite{de2009} or naive  persistence approaches typically have small computational cost. 
But they are often outperformed by more sophisticated approaches like Support Vector Regression (SVR) \cite{guo2006} and Artificial Neural Networks (ANNs) such as Long Short Term Memory (LSTM) \cite{kong2017, brucke2021}. However, these are oftentimes computationally very expensive. 
A recent study showed the application of different Reservoir Computing (RC) techniques for energy demand forecasting in the building context with high forecast quality and small computational effort \cite{brucke2024benchmarking}. 
Besides computational cost and forecasting quality, the need of expert knowledge or required amounts of data are additional criteria for evaluating the methodologies. Forecasting models with different characteristics are often competing against each other but the model selection procedure tends to be complicated multi-criteria decision process. Nevertheless, it is mostly carried out ex-post by choosing the model with smallest forecasting error \cite{simani2024, sun2020, guo2018}. However, energy management in real world systems like buildings or electric vehicles is performed in real time \cite{kwak2015, guo2021, quan2021}. This emphasizes the need for continuous real-time model selection of competing forecast techniques. 
Additionally, it is important to be able to quantify and limit the risk of decision making for one or the other forecasting model in real world energy management.
Otherwise, a false decision - especially with unknown decision risk - could have negative economic implications for the different stakeholders like distribution grid operators or the building owners.
In the following subsection, a brief overview on existing model selection methods for competing forecasters is outlined. 

\subsection{Model selection methodologies}
\label{sec:literature}
Compared to  research on new forecasting methodologies which is extensive throughout different domains of application, research on model selection approaches for competing forecasts has not yet gained a similar level of attention. 
This is particularly apparent in the field of applied research, and -- in the perception of many applied scientists -- has been lacking clearcut principled guidelines to some extent \cite{practical_guide_model_selection}.  
As already mentioned above, most publications that perform model selection in the energy context compare error measures ex-post on a given test data set which can be considered a naive benchmark method. 
However, due to the limited data availability at a time $t$ in the operation, this approach is not really applicable for real time model selection. 
To overcome this, the authors in \cite{swanson1997} use out-of-sample forecast-based model selection criteria for real time macroeconomic forecasting. 
The Akaike information criterion (AIC) and the Bayes information criterion (BIC) are also often used for model selection as in \cite{billah2006}. 
Besides the AIC, the authors of \cite{liu2022} additionally use the Pearson and Spearman correlation index for model selection. 

Given the stochastic nature of the observations underlying model selection a thorough approach would aim to control the stochastic uncertainty by means of probabilistic guarantees such as specifying a significance or confidence level $\alpha$. 
In the model selection context, such a guarantee means to select the forecast model with smaller error based on past average performance of a given time period. 

Traditionally, these guarantees relied on strong structural assumptions on the distribution of the observations, such as stationarity, ergodicity and typically along with assumptions controlling the decay of dependence over time such as mixing conditions or weak dependence conditions as laid out in great generality in e.g. \cite{douk}.

In energy forecasting, though, typical seasonal effects and intraday patterns indicate that stationarity cannot hold for the original observations. 
Preprocessing techniques like detrending and deseasonalization can mitigate some of these violations, but  other assumptions, such as ergodicity or weak dependence, remain challenging to rigorously test. 

Moreover, we are  not heading for only one model selection decision \glqq once for all\grqq~to be taken retrospectively and to be based on a fixed set of observations, but rather for continuous model selection in the same rhythm as the incoming new observations, which can be seen as a sequential decision approach as brought up in Wald's seminal  paper \cite{wald}. 
Such approaches inherently must consider information simultaneously along a path of observations which seems to suggest the need of even stronger structural assumptions, thus narrowing the application scope. 
Nevertheless, in particular settings, such continuous-time decision procedures have found their way to practical applications, as in the well known CUSUM-type control charts (e.g., \cite[Figure 2]{Cusum}) of stochastic quality control. However, when applied to dynamic settings with time dependent observations, as in energy forecasting, the structural assumptions mentioned above traditionally could not be dispensed with.

With the recent advent of many new powerful procedures based on so-called \glqq e-values\grqq, see~\cite{gametheoretic,TestMartingalesBayesFactors}, the necessity of these structural assumptions could almost entirely be dropped. 
Methods based on $e$-values are now also covering arbitrarily non-stationary situations in time-dynamic settings, therefore enable continuous or real time model selection of competing forecasting models.
Additionally, $e$-values provide strong probabilistic guarantees, which was out of scope for standard machine learning tools.  
Methods based on $e$-values were brought to continuous-time model selection in \cite{valid_sequential_inference_on_probability_forecast_performance,choe2023comparing}. 
Technically, such methods amount to control whether a certain evidence measure remains within or crosses certain (simultaneous) control bounds much the same way as in the mentioned control charts of CUSUM-type. 

Despite the obvious advantages of continuous-time model selection, $e$-values so far have not been applied in the context of building energy management to the best of our knowledge. In addition, so far $e$-values-based model selection was limited to predicting binary outcomes. So in this sense extending the applicability of $e$-values-based CUSUM-like charts for forecast model selection to prediction of continuous outcomes like energy demand should be welcome.\\
A well established measure for forecast accuracy in the context of
building energy management is the mean absolute error (MAE), see
\cite{mae_load_forecasts}. So if possible such an $e$-value-based approach should use MAE as a score within its decision finding.

\subsection{Contribution of this study}
In this work, we present the application of $e$-values for real time  model selection in the field of building energy demand forecasts. 
For that, we use the results from a recent study \cite{brucke2024benchmarking} where the authors compare and benchmark different recurrent network based forecasting methodologies. 
Their models result in forecasts with varying characteristics. 
While the LSTM approach yields small mean absolute errors (MAEs) with very smooth demand forecasts, a method called Next-Generation-Reservoir-Computing (NG-RC) is able to predict the erratic behavior of energy demands with slightly higher MAEs on average. 
These different characteristics make it hard to decide for one model for the whole period of time. 
Therefore, we carry out a continuous model selection procedure for electricity demand forecasts based on e-values with a fixed decision risk of 5\% being guaranteed. 
We benchmark our decision making procedure against multiple persistence approaches for model selection. 
Furthermore, we propose and discuss several strategies for the case when the evaluated $e$-values do not favor one of the considered forecasting models with significance. 
Of course, the warranty given by the e-values comes with a price and we cannot exclude that other decision tools might in some scenarios show better forecast performance, but then, based on the same available information, they cannot offer level alpha control.
Albeit this price, in the sequel we will show that our procedure can be tuned to be competitive.
\medskip

This paper is structured as follows. First, in Section~\ref{sec:methodology}, we begin by introducing the mathematical concept of $e$-values, along with relevant definitions and notations from existing literature. We then detail the construction of a sequential hypothesis test, which also leads to the development of confidence sequences for a given significance level. Building on this foundation, we introduce the $e$-selection procedure, which extends the sequential test to generate new predictions by dynamically combining the compared forecasting methods.
Afterwards, in Section ~\ref{data description}, we present the electricity demand time series data used in our application, along with the forecasting techniques employed. We also discuss data transformation methods and provide benchmark scores for comparison. In Section~\ref{sec:results} we present the results of our study, including the performance of the $e$-selection procedure and an analysis of computational runtime. Finally, in Section~\ref{sec:discussion}, Finally, we discuss the limitations of our proposed procedure and provide further insights into the guarantees it offers. We also offer a brief outlook on potential directions for future research.\\

%% file: Methodology.tex
In this section, we briefly introduce the mathematical theory of $e$-processes, which provide a framework for so-called \textit{Sequential Anytime-Valid Inference} (SAVI). SAVI refers to statistical inference (i.e., tests, confidence bands, decisions) along with probabilistic error control in a sequential (possibly even continuous-time) setting, often without relying on distributional assumptions of the observed entities (in our case the electricity demands) \cite{gametheoretic, SafeTesting, ramdas2022admissible}. 
The \textit{``Anytime-Valid''} in SAVI alludes to the key feature that $e$-processes allow for \textit{optional stopping}, which is essential for the task to choose between two candidate forecast procedures in a continuous time setting under probabilistic error control as in 
\cite{valid_sequential_inference_on_probability_forecast_performance}. Here, \textit{optional stopping} refers to the fact that we monitor an evidence measure continuously over time along with the incoming observations, and \textit{optionally} stop and make a decision, at a random time that is not known when monitoring starts, when sufficient evidence has accumulated so that the probability for a false decision can be kept controlled. In the context of continuous-time model selection, this decision will determine which forecast is superior to the other.\\

We begin by formally introducing $e$-values in Subsection~\ref{fundamentals}. Subsequently, we set up corresponding test statistics based on scores for the competing forecasts in Subsection~\ref{Setting}. These test statistics are used to distinguish formal hypotheses introduced in Subsection~\ref{Hypoth} leading to an $e$-selection procedure in Subsection~\ref{procedure}.

\subsection{Fundamentals for \textit{e}-Values and Their Relations to \textit{p}-Values}\label{fundamentals}

An $e$-process $(E_t)$ is a nonnegative stochastic process\footnote{For our purposes, a stochastic process is a stream of possibly time-dependent random variables indexed by their observation time $t$, and $t$ may range in an ordered set of time points $\mathcal{T}$, $\mathcal{T}$ discrete or continuous. \cite{probability_theory}} whose expected value $\E_P[E_\tau]$ is upper bounded by one for any arbitrary stopping time\footnote{A stopping time is a random time $\tau$, for which for each time $t$, the accumulated information ${\cal F}_t$ at time $t$ is sufficient to decide whether $\tau\leq t$. \cite{probability_theory}} $\tau$ under a given null hypothesis $\mathcal{H}_0$ \cite{gametheoretic}. Formally, this is expressed as:
    \begin{equation} \label{E lt 1}
        \E_P[E_\tau] \leq 1 \qquad \text{ for all stopping times } \tau  \text{ and } P \in \mathcal{H}_0. 
    \end{equation}
Apart from Eq.~\eqref{E lt 1} no further assumptions on the distribution or on the (stochastic) dependence of $(E_t)$ over time are made, which
makes $e$-processes particularly appealing for our time-dependent, non-stationary sequences of electricity demands.

A realization of an $e$-process is called $e$-value. This concept originates from the term \glqq betting score\grqq~by Shafer in  \cite{betting_score}.  
While a $p$-value represents the probability, under the assumption the null hypothesis holds, to observe, in a new experiment, a value for the test statistic being at least as large as the one obeserved, an $e$-value measures the accumulation of evidence against this hypothesis, growing rapidly when the hypothesis is violated \cite{valid_sequential_inference_on_probability_forecast_performance, gametheoretic}. 
As expectations, $e$-values can simply be combined by averaging them, with 
the average remaining an $e$-value, which is an important advantage over $p$-values in a sequential setting, compare  \cite{vovk2021}. 
The relevance of $e$-values for testing lends to Ville´s inequality which entails the following bound valid for any $e$-process $(E_t)$
\begin{equation}\label{fdr}
        P(E_\tau \geq 1/\alpha) \leq \alpha~ \text{ for all stopping times } \tau  \text{ and } P \in \mathcal{H}_0.
 \end{equation}
By means of Eq.~\eqref{fdr}, any $e$-process $(E_t)$ can be translated into a sequential hypothesis test controlling the (familywise\footnote{The \textit{familywise error rate} (FWER) is a key concept in multiple testing / simultaneous inference, compare \cite{MultipleTesting}.
It denotes the probability, under $\mathcal{H}_0$ to falsely reject at least one of the multiple hypotheses; this corresponds to simultaneous confidence intervals giving a probabilistic guarantee that the intervals \textit{simultaneously} cover the unknown parameter or outcome with a given level. FWER control implies the weaker form of \textit{false discovery rate} (FDR) control which only warrants that the ratio of falsely rejected hypotheses stays below a given bound.}) type-I error for a given null hypothesis
$\mathcal{H}_0$, compare \cite{gametheoretic,SafeTesting, HowardEtAlTimeUnifConf,howard2020timeuniform}.
More specifically, we can reject the null hypothesis $\mathcal{H}_0$ at a familywise significance level of $\alpha$ as soon as $E_\tau$ surpasses the value $1/\alpha$ from below.
While providing anytime-valid inference \cite{gametheoretic, valid_sequential_inference_on_probability_forecast_performance}, $e$-values-based methods typically result in lower statistical power compared to those designed for fixed sample sizes (often called pointwise), but of course the pointwise error guarantee then only is valid  for one time point, and combining pointwise guarantees for several time points bears the risk of alpha error cumulation \cite{multiple_testing}.
The connection to $p$-values is given by the fact that by taking 
 \begin{equation}\label{e_to_p_value}
     \mathfrak{p}_\tau = \min(1, 1/E_{\tau}),
 \end{equation}
any $e$-value can be converted into a conservative $p$-value \cite{vovk2021}. 

As indicated, these $e$- and $p$-values are used to make decisions about which forecaster to prefer. To do so, we formally introduce the competing forecasts, the scores assessing their accuracy and a (sequence of) test statistics based on the differences of these scores. In this dynamic setting we must also carefully specify the amount of information available for the test statistics at a given time instant $t$. Much of this layout closely follows \cite{choe2023comparing} who consider continuous time superiority testing for binary predictions. In fact, one contribution of ours is to extend their setting to superiority decisions for unbounded sores such as the mean absolute error (MAE).

\subsection{Definitions and Formalizations: Information Sets, Scoring Rules, and Test Statistics}\label{Setting}
Following the formulations in \cite{choe2023comparing}, we start with two procedures issuing forecasts, denoted as $(p_t)$, $(q_t)$, indexed by a time index  $t\in \{1,...,T\}\subset \N$, along with the corresponding outcomes $(y_t)$ generated from the real distributions $(r_t)$.
We make no assumptions on the sample rate, or require equidistant observations over time.  To simplify notation, we drop the interpretation of $t$ as a unit of time. Instead, $t$ serves as a counting index for the considered predictions, and the time span between $t-1$ and $t$ is not used in the analysis. To specify the available information at time $t$ for a specific process, we use the filtrations defined in \cite{choe2023comparing}. For example,   $\mathcal{F}_t$ represents the \textit{oracle} filtration knowing all information up to time $t$.

The forecasts are evaluated using a score $S$, more specifically, in our case 
of the MAE, $S(x,y)=|x-y|$ where $x$ is the value of the forecast and $y$ is the actual outcome. 
We are well aware of the fact, that in the context of binary decisions, this function $S$, read as a scoring rule, is not proper in the terminology of \cite{strictly_proper}, as required in \cite{choe2023comparing}. However, this does not negatively impact our application, which deals with continuous outcomes.

The evidence for a specific forecast is evaluated through the empirical score differences, defined as 
\begin{equation}\label{score_diff}
    \widehat{\delta}_t := S(p_t,y_t) - S(q_t,y_t), 
\end{equation}
which comes with a forecast counterpart 
\begin{equation*}
\delta_t := \E[S(p_t, y_t) - S(q_t, y_t)\vert \mathcal{F}_{t-1}],
\end{equation*}
taking into account the conditional expectation given the information up to time $t-1$. Following \cite{choe2023comparing}, instead of focusing on the the varying sequence $\widehat{\delta}_t$, we want to base our decisions on the average empirical score differentials
\begin{equation}\label{average_score_diff}
    \widehat{\Delta}_t := \frac{1}{t} \sum_{i=1}^t\widehat{\delta}_i
\end{equation}
together with the unobservable sequence of the average expected score differentials $\Delta_t=\frac{1}{t}\sum_{i=1}^t \delta_i$. 
The sequence $\Delta_t$ indicates whether one forecast outperforms the other one on average. Since the MAE is a negatively oriented scoring rule, negative score differentials indicate a perference for forecast $(p_t)$. 

Now in the approach in \cite{choe2023comparing}
and also in most techniques discussed in \cite{HowardEtAlTimeUnifConf}, for powerful tests and hence efficient model selection, it is crucial to be able to use some exponential concentration bounds constructed in similar manour as  the Hoeffding or the Bernstein inequality. To this end, we require that the empirical score differentials $\widehat{\delta}_t$ should be bounded, meaning we assume 
\begin{equation}\label{bounded_score}
    \vert \widehat{\delta}_t \vert \leq \frac{1}{2} \text{ for all } t\geq 1.\footnote{
    In the original reference, it is written as $\vert \widehat{\delta}_t \vert \leq c/2 \text{ for some } c\in (0,\infty)$, but we use $c=1$ and adapt the equations accordingly.
    }
\end{equation}
Eq.~\eqref{bounded_score} in general is violated for the MAE. 
To address this issue, we obtain bounded scores by transforming our empirical score differentials $\widehat{\delta}$ using an appropriate sigmoid function $f$ in Section~\ref{data_preprocessing}.

To construct a suitable $e$-process, we follow the approach of \cite{choe2023comparing} based on \cite{howard2020timeuniform}, which involves using a cumulative sum process $M_t$, whose deviations from $0$ we want to control over time.
In our case $M_t=\sum_{i=1}^t \hat \delta_i-\delta_i$, respectively, or after transformation $M_t=\sum_{i=1}^t \tilde \delta_i-\E[\tilde \delta_i\,|\,\mathcal{F}_{i-1}]$. 
From now on,
we only work with the transformed score differences and drop the notational distinction between the transformed and untransformed score differences and for better reference rather use the notation for the untransformed ones taken from \cite{choe2023comparing,howard2020timeuniform}.

\begin{figure}[H]
    \centering
    \begin{subfigure}[t]{0.315\textwidth}
        \centering
        \includegraphics[width=\textwidth]{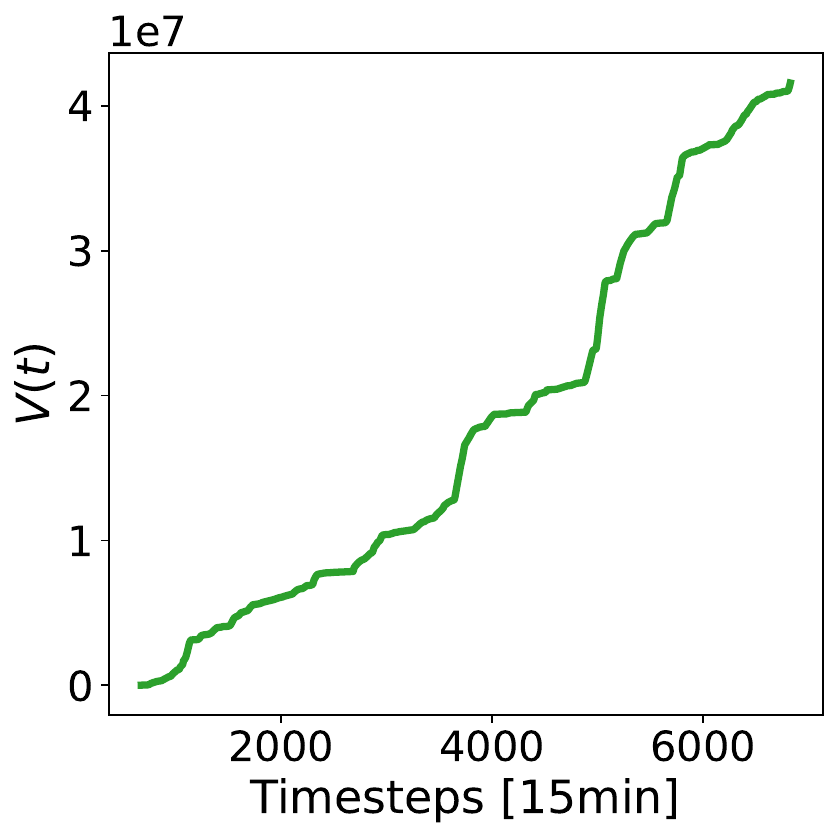}
        \label{var_process}
    \end{subfigure}
    \hspace{2cm}
    \begin{subfigure}[t]{0.3\textwidth}
        \flushright
    \includegraphics[width=\textwidth]{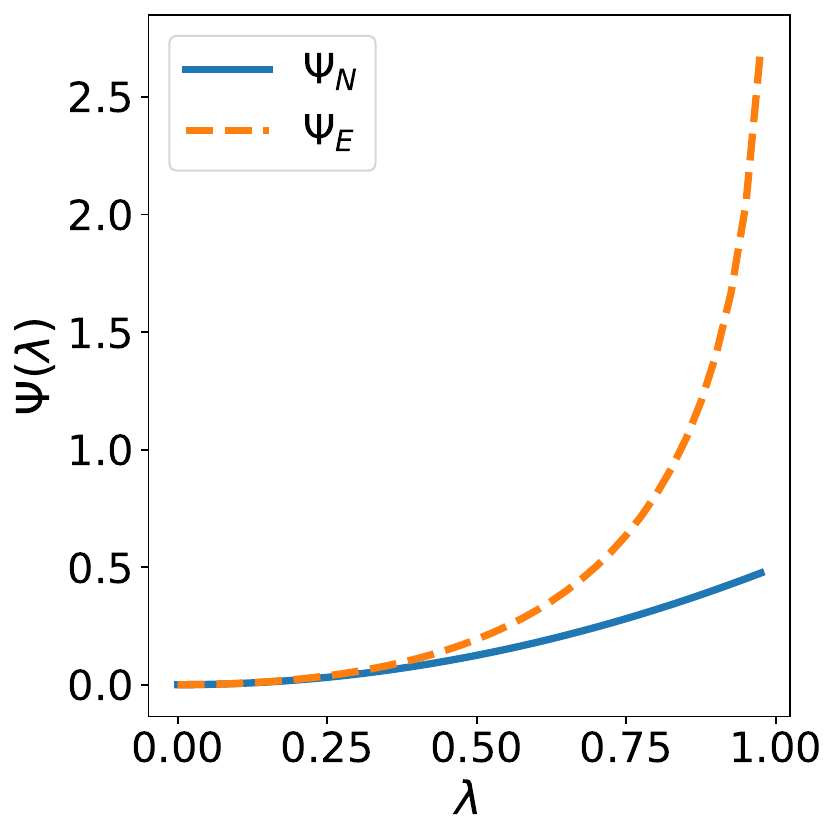}
        \label{psi_function}
    \end{subfigure}
    \caption{Visualization of the required functions from Eq.~\eqref{e-process}. \textbf{(Left)} Illustration of the variance process $\widehat{V}_t$ using the data described in Section~\ref{data description}. \textbf{(Right)} Sub-exponential and sub-gaussian $\psi$-functions $\psi_E$ and $\psi_N$ for different $\lambda \in [0,1)$. }
    \label{psi_var}
\end{figure}

A natural way to generate an $e$-process  is through the exponential transform\footnote{
Mathematically, if $M_t$ forms a martingale, the exponential transform $e^{\lambda M_t}$ results in a submartingale due to Jensen's inequality. This submartingale can be transformed into a supermartingale $e^{\lambda M_t - \psi(\lambda) V_t}$ for appropriate choices of $\psi$ and $V_t$. By the definition of this supermartingale, $E_\lambda(t)$ forms an $e$-process \cite{howard2020timeuniform, HowardEtAlTimeUnifConf, probability_theory}.
}
\begin{equation}\label{e-process}
    E_{\lambda}(t) = \exp\left(\lambda M_t - \psi(\lambda) V_t\right),
\end{equation}
with appropriate choices for $V_t, ~\lambda, ~ \psi(\lambda)$.
In this formulation
\begin{itemize}
     \item $V_t$ is the variance process for $M_t$ and can be interpreted as a measure of intrinsic time to quantify the deviations of $M_t$ from zero. In our setting, we adopt the default variance process defined by \cite{choe2023comparing}, which is given by:
     \begin{equation}
    \hat V_t =\sum_{i=1}^t (\hat \delta_i-\hat\Delta_{i-1})^2, 
    \end{equation}
    where $\widehat{\delta}_i$ and $\widehat{\Delta}_{i-1}$ are defined in Eq.~\eqref{score_diff} and Eq.~\eqref{average_score_diff}.
    An illustrative example of the process $\widehat{V}_t$ is shown in   Figure~\ref{psi_var} (left).

    \item $\lambda > 0$ is a hyperparameter that controls the growth rate of the $e$-process $E_{\lambda} (t)$. It adjusts the impact of the deviations $M_t$ by assigning them a weight. Specifically, larger values of $\lambda$ increase the weights of extreme deviations. Therefore, $\lambda$ also controls the willingness to take risk, compare \cite{HowardEtAlTimeUnifConf, howard2020timeuniform}.
    
    \item $\psi(\lambda)$ is a
 cumulant-generating like function (CGF-like), that determines the rate at which the process $M_t$ can grow in relation to the intrinsic time $V_t$. In \cite{howard2020timeuniform}, several useful $\psi$-functions are discussed, but we focus on the sub-exponential function given by:
 \begin{equation}
     \psi_E(\lambda)= -\log(1-\lambda)-\lambda \text{ for all } \lambda \in [0,1).
 \end{equation}
 In Figure~\ref{psi_var} (right) are shown the sub-exponential $\psi$-function $\psi_E$ along with the sub-gaussian function $\psi_N$ which is described in  \cite{choe2023comparing, howard2020timeuniform}. 
\end{itemize}    
     
\subsection{Sequential Tests using \textit{e}-Processes} \label{Hypoth}
Our goal is to sequentially test if one forecast outperforms the other on average, with a given significance level $\alpha$. This means we want to test whether the sequence $\Delta_t$ is positive or negative for all timesteps $t$. Formally, the null hypothesis is defined as 
\begin{equation}\label{test_p}
    \mathcal{H}_0(p,q):~ \Delta_t \leq 0 \text{ for all times } t\geq 1.
\end{equation}
For a negatively oriented scoring rule like the MAE, the hypothesis implies that the forecast $(q_t)$ is not better than the forecast $(p_t)$ on average across all times $t$. Analogously, by switching the order of the forecasts, we get
\begin{equation}\label{test_q}
     \mathcal{H}_0(q,p):~ \Delta_t \geq 0 \text{ for all times } t\geq 1.
\end{equation}
The complete sequential test is then  constructed by combining the two separate tests defined in \eqref{test_p} and \eqref{test_q}, each with a significance level of $\alpha/2$. In our approach, we utilize Theorem 3 from \cite{choe2023comparing} as the main result. Accordingly the $e$-process for testing $\mathcal{H}_0(p,q)$ is defined as
\begin{equation}\label{e-process_thm}
    E_{\lambda}(t) := \exp\left(\lambda t \widehat{\Delta}_t - \psi_{E}(\lambda) \widehat{V}_t \right) \text{ for } \lambda \in[0,1).
\end{equation}
We write $E^*_\lambda (t)$ as the $e$-process for testing $\mathcal{H}_0(q,p)$. \textbf{Therefore the hypotheses $\mathcal{H}_0(p,q)$ or $\mathcal{H}_0(q,p)$ are rejected, if the corresponding $e$-process passes the threshold of $\alpha/2$ (Eq~\eqref{fdr}) from below.}\\

This construction of the $e$-process $E_{\lambda}(t)$ , comes with time-uniform confidence sequences for $\Delta_t$, which provide coverage guarantees that hold uniformly over time, ensuring that the process remains within the confidence interval at all times. This allows us to reject the hypotheses if the entire confidence sequence lies completely above or below zero \cite{choe2023comparing}.
Rewriting Eq.~\eqref{fdr} and using the $e$-process from Eq.~\eqref{e-process_thm}, we obtain the uniform boundary 
\begin{equation}
    u_{\alpha/2} = \frac{\psi(\lambda) \widehat{V}_t - \log(\alpha/2)}{\lambda},
\end{equation}
which then leads to the symmetric $(1-\alpha)$-confidence sequence for $\Delta_t$:
    \begin{equation}\label{eq:confidence_interval}
        C_{\alpha}(t) = \left[\widehat{\Delta}_t - \frac{u_{\alpha/2}}{t}; \widehat{\Delta}_t + \frac{u_{\alpha/2}}{t} \right],
    \end{equation}
    such that 
    \begin{equation}
       \P(\Delta_t \in C_{\alpha}(t)) \geq 1-\alpha ~ \text{ for all } t\geq 1. 
    \end{equation}
This formulation ensures that the process $\Delta_t$ remains within the confidence sequence with a probability of at least $1-\alpha$ simultaneously for all time steps $t$.\\

A plot of $\widehat \delta_t$ (or $\widehat{\Delta}_t$) along with the confidence sequence $C_{\alpha}(t)$ over time can be seen as a variant of a CUSUM-type control chart, allowing for similar interpretation. Specifically, crossings of the $C_{\alpha}(t)$ bounds indicate critical events. For easier interpretation,  we backtransform the
score differences and differentials as well as the upper and lower confidence bounds to
the original MAE-scale, which in our application is measured in Watts. An example of such a chart is shown in Figure~\ref{fig:processes} (left). 

\subsection{\textit{e}-Selection Procedure}\label{procedure}

In this section, we outline the procedure for selecting the forecasting method for future time steps using sequential tests, which we refer to as \glqq$e$-selection\grqq. The goal is to construct a new prediction at each time step by combining two existing forecasting methods. The procedure can be roughly divided into three steps, which are performed at every time step $t$. The starting point is the empirical score differentials $\widehat{\delta}_t$ from Eq.~\eqref{score_diff}, along with a predefined significance level $\alpha \in (0,1)$. The procedure involves calculating the $e$-processes, conducting the corresponding sequential tests as a descriptive task, and combining the forecasting methods to generate a new prediction. The steps are summarized as follows:

\begin{enumerate}
    \item[1.] \textit{Calculate $e$-Processes:}\\
First, we need to select the hyperparameter $\lambda$ and the additional parameter $\omega$, which represents the number of past observations to consider. Instead of accounting for the entire time horizon of the forecasting methods, we focus only on the most recent $\omega$ observations to better capture the dynamic behavior. Therefore the expresssion $\widehat{\Delta}_t$ in Eq.~\eqref{average_score_diff} is changed to rolling average
\begin{equation}
    \widehat{\Delta}_{t,\omega} = \frac{1}{\omega} \sum_{i=t-\omega}^t \widehat{\delta}_i \text{ for } t\geq \omega,
\end{equation}
and analogously $\widehat{V}_{t,\omega}$ and $E_{\lambda, \omega}(t)$ with the $e$-process started at time step $t-\omega$.
By allowing optional stopping and continuation, the $e$-processes remain valid under this approach \cite{valid_sequential_inference_on_probability_forecast_performance}. The choice of $\lambda$ and $\omega$ will be discussed in Section~\ref{sec:results}. To obtain bounded score differentials $\widetilde{\delta}_t$ we use the transformation described in Section~\ref{data_preprocessing}. We then use the values $E_{\lambda, \omega}$ and $E^*_{\lambda, \omega}$ at time step $t$ for the sequential test. \\

\item[2.] \textit{Sequential Test:}\\
For the test, we simply check whether the values $E_{\lambda, \omega}$ and $E^*_{\lambda, \omega}$ exceed the threshold of $\alpha/2$. We reject $\mathcal{H}_0(p,q)$, if $ E_{\lambda, \omega} \geq \alpha/2 $, which indicates that forecast $(q_t)$ outperforms forecast $(p_t)$ on average over this specific period. Similarly, we reject  $\mathcal{H}_0(q,p)$, if $ E^*_{\lambda, \omega} \geq \alpha/2$. In cases where neither hypothesis can be rejected, there is insufficient evidence to favor one forecast over the other, and we will discuss the subsequent steps in the following section.\\

\item[3.] \textit{Generate Predictions by Combining Forecasting Methods (Forecast Fusion):}\\
Based on the results of the sequential test, we want to generate a combined prediction $(\theta_t)$, termed forecast fusion \cite{fusion}. While the test itself does not rely on distributional assumptions, we do need some assumptions regarding future performance \cite{choe2023comparing}. A simple approach is to use a persistence model, where if forecast $(p_t)$ outperformed $(q_t)$ on the previous day, we will also select forecast $(p_t)$ for the next day. This ensures that $(\theta_t)$ is selected based on information available up to time $t-96$, assuming that the forecasts $(p_t)$ and $(q_t)$ are also $\mathcal{F}_{t-96}$-measurable. However, there are time steps where neither hypothesis can be rejected. For these cases, a different selection approach is necessary.
Formally, the prediction at time step $t$ can be expressed as: 
\begin{equation}
    \theta_t = \begin{cases}
                p_t & \text{ if } E^*_{\lambda, \omega }(t-96) \geq \alpha/2\\
                q_t & \text{ if } E_{\lambda, \omega }(t-96) \geq \alpha/2\\
                z_t & \text{ else}, 
                \end{cases}
\end{equation}
where $z_t$ is determined using one of the following three approaches:
\begin{itemize}
    \item[i)] \textit{Persistence:} The simplest option is the persistence $e$-selection. Once a decision is made at a time step $s<t$, an appropriate strategy is to continue using that selected forecasting method as long as there is insufficient evidence to switch. In this case,
    \begin{equation}
        z_t = \theta_{t-1}.
    \end{equation}
  The advantage of this method is that it minimizes the number of switches between the possible predictions $(p_t)$ and $(q_t)$, leading to a more stable forecasting process.
    \item[ii)] \textit{Sampling:} The second strategy involves randomly sampling of the forecast methods based on the amount of evidence, as indicated by the corresponding $e$-value. In this approach, we calculate the weight that determines the probability of selecting each method. Therefore, we use Eq.~\eqref{e_to_p_value} to transform the $e$-values into anytime-valid $p$-values. We write:
    \begin{align*}
        \mathfrak{p}_t &= \min(1, 1/E_{\lambda, \omega}(t)) \\
        \mathfrak{p}^*_t &= \min(1, 1/E^*_{\lambda, \omega}(t)), 
    \end{align*}
    which results in the sampling weight for the forecast $(p_t)$ as:
    \begin{equation}
        w_{p,t} = \frac{1 + \mathfrak{p}_t - \mathfrak{p}^*_t}{2}
    \end{equation}
    and similarly, the weight for forecast $(q_t)$ as:
    \begin{equation}
        w_{q,t} = \frac{1 + \mathfrak{p}^*_t - \mathfrak{p}_t}{2} = 1-w_{p,t}.
    \end{equation}
    In this case 
    \begin{equation}
        z_t = \begin{cases}
            p_t \text{ with probability } w_{p,t}\\
            q_t \text{ with probability } w_{q,t}.
        \end{cases}
    \end{equation}
    When $E_{\lambda,\omega}(t)$ and $E^*_{\lambda,\omega}(t)$ are both less than 1, indicating that there is no sufficient evidence to reject either hypothesis, the weights reduce to 1/2 for each forecast. This results in a 50/50 chance of selecting either forecast $p_t$ or $q_t$. 
    \item[iii)] \textit{Weighted average (wAvg):} The weighted average approach combines the forecasts $(p_t)$ and $(q_t)$ based on the weight $w_{p,t}$ and $w_{q,t}$. Therefore
    \begin{equation}
        z_t = w_{p,t} \cdot p_t + w_{q,t} \cdot q_t.
    \end{equation}

As is well known and can be easily  proven, the sampling strategy~ii) of selecting randomly among the forecasts is, in expectation, equivalent to the weighted average approach~iii). However, the variance of approach~iii) is strictly lower than that of approach~ii), unless either $w_{p,t} \in \{0, 1\}$ or $p_t = q_t$ with probability 1. Therefore, when combining two forecasts at the same time $t$, the weighted approach generally provides increased statistical power. Still, approach~ii) will be advisable, if constraints such as budget limitations  preclude approach~iii), limiting,the number of forecasters to be used per time to just one.
\end{itemize}
\end{enumerate}
After outlining our procedure for generating predictions by combining two forecasting methods, we now turn our attention to the application of this approach. Specifically, we apply our procedure to forecast electricity demand using time series data. In the following section, we introduce the electricity demand time series data and the forecasting techniques employed, which together form the basis of our application case.

%% file: data.tex
To show the application of e-values for the real-time forecast comparison in the building context the results from a recent publication \cite{brucke2024benchmarking} are used. 
In the cited reference, the authors compare different forecasting methodologies for household electricity demand forecasting. 
In this work, we use their forecasting results, namely for the methods \textit{Long-Short-Term-Memory} (LSTM)  and \textit{Next Generation Reservoir Computing} (NG-RC) which are briefly introduced in Section~\ref{lstm} and Section~\ref{ngrc}. Section~\ref{data basis} shortly describes the raw data that was used to obtain the forecasts. 

\subsection{Raw Data}\label{data basis}
The electricity demand time series used for forecasting are derived from the \textit{EMS3} data set within the \textit{EMSIG} data set \cite{dataset}. This data set represents the energy data recorded by a decentralized household energy management systems (EMS) from the DACH region, with a 15 minutes resolution in Watts [W]. Day-ahead predictions are created for the sum of the active power of electricty consumption which is denoted by the column \texttt{sum\_consumption\_active\_power} in the EMS3 data set. The models were trained and validated using the first 90\% of the data set, covering the period from January 1, 2019 to October 21, 2020, 21:00. The test set, which will be used for real-time forecasting comparisons,
consists of the remaining 10\% of the data points, spanning from October 21, 2020, 21:15, to December 31, 2020 and includes a total of $N=6826$ measured electricity demand data points. \cite{brucke2024benchmarking} 

\subsection{Next Generation Reservoir Computing}\label{ngrc}
Next Generation Reservoir Computing (NG-RC) is a machine learning algorithm that originates from nonlinear vector autoregression (NVAR), designed for analyzing  dynamical chaotic systems using observed time-series information \cite{ngrc_paper, ng_rc_paper2}. 
Unlike traditional reservoir computing, NG-RC constructs its feature vectors directly from unique polynomials of time-delayed input signals. 
This approach requires only a small amount of training data and yields computationally inexpensive optimization which results in a highly efficient algorithm as is describes in more detail in \cite{brucke2024benchmarking}.

\subsection{Long-Short-Term-Memory Neural Networks}\label{lstm}
Long Short-Term Memory neural networks (LSTM), introduced by Hochreiter and Schmidhuber, are a special form of recurrent neural networks (RNNs) designed to handle the vanishing gradient problem of conventional RNNs. LSTMs fit very well for processing sequential data as they effectively capture long-term dependencies while retaining the ability to recognise short-term patterns \cite{lstm}. 
This ability results from three categories of gates for each memory cell: Input, output and forget gates. These gates regulate the storage and discarding of information and ensure that relevant data is retained over long sequences. Therefore, LSTMs are widely recognised as state-of-the-art for tasks such as time series prediction. 
\cite{brucke2024benchmarking} \\

\subsection{Data Preprocessing}\label{data_preprocessing}
\begin{figure}[H]
    \centering
    \includegraphics[width = \textwidth]{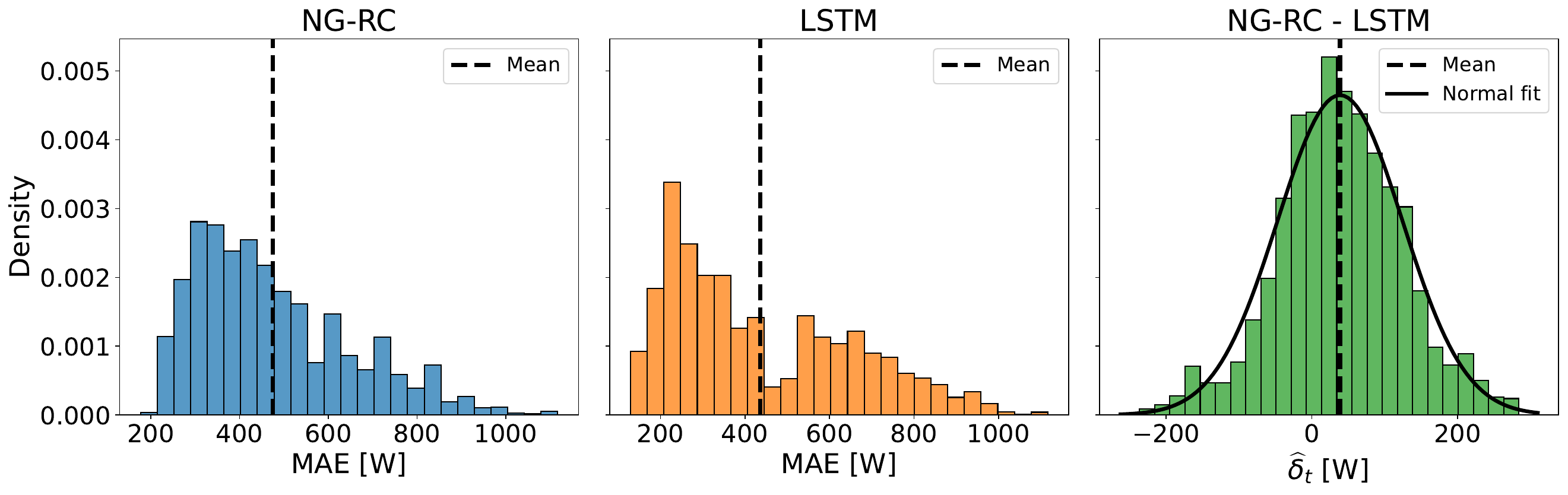}
    \caption{Histograms of the MAE of NG-RC (left), LSTM (middle) and the empirical score differences $\widehat{\delta_t}$ (right) with mean values (dotted line) in watts, and normal fit (solid line).}
    \label{hist_scores}
\end{figure}

Each model generates a prediction for every time step of the test set for the following 24 hours, resulting in a prediction matrix $\widehat{y}=(\widehat{y}_{tk})$, where $t\in\{1,..., 6826\}$ denotes the time step and $k\in \{1,...,96\}$ denotes $k$-step-ahead prediction horizon. The real power consumption is denoted by $y=(y_{tk})$ with the same dimension as $\widehat{y}$. Every row in $y$ contains the measured power consumption of the respective household from time step $t$ for the next 24 hours. 
Accordingly, from one row $t$ to the next row $t+1$, the data is shifted by one time step.  
We then use the MAE as a standard metric for evaluating load forecasts \cite{load_forecast_review}. 
The MAE for each time step $t$ is calculated using the following equation:  
\begin{equation}\label{MAE}
    \text{MAE}(\widehat{y}_t, y_t) = \frac{1}{96} \sum_{k=1}^{96} \vert \widehat{y}_{tk} - y_{tk} \vert,  
\end{equation}
where $y_{tk}$ denoting the true realized value at time $t+k$.
Let $(p_{tk})$ represent the prediction from NG-RC and $(q_{tk})$ the prediction from LSTM. 
We calculate the empirical score differences $\widehat{\delta}_t$ between NG-RC and LSTM as follows: 
\begin{equation*}
    \widehat{\delta}_t =\text{MAE}(p_t, y_t) -\text{MAE}(q_t, y_t).  
\end{equation*}
Figure~\ref{hist_scores} shows the histogram of the MAE scores for NG-RC (left) and LSTM (middle) along with the score differences $\widehat{\delta}_t$ on the right hand side of the figure. 
The overall mean value is shown by a dashed vertical line in every histogram. The score differences can be approximated by a shifted normal distribution.

To apply the procedure from Section~\ref{procedure} we need bounded scores like in Eq.~\eqref{bounded_score}, which can be obtained using an appropriate transformation function $f(x)$ which is chosen to be a sigmoid function in this work.
More specifically, we require $f$ to satisfy the following near-to-canonical
conditions: 
\begin{itemize}
    \item[(i)] To minimize squeezing effects, $f$ should be approximately linear in the central region where most of the probability mass is concentrated.
    \item[(ii)]  $f$ should exhibit odd symmetry, with $f(0)=0$.
    \item[(iii)]  To ensure boundedness, $f$ should be curved in the tails.
\end{itemize}

The actual choice of $f$ according (i)--(iii) is of secondary importance and different such choices will only lead to minor differences in the results. One possible such choice is
\begin{equation}\label{sigmoid_eq}
    f(x) = \Phi(x/\sigma) - 1/2,
\end{equation}
where, $\Phi(x)$ is the cumulative distribution function (CDF) of the standard normal distribution and $\sigma$ is an appropriate scaling parameter.

\begin{figure}[H]
    \centering
    \begin{subfigure}[b]{0.4\textwidth}
        \flushright
        \includegraphics[width=\textwidth]{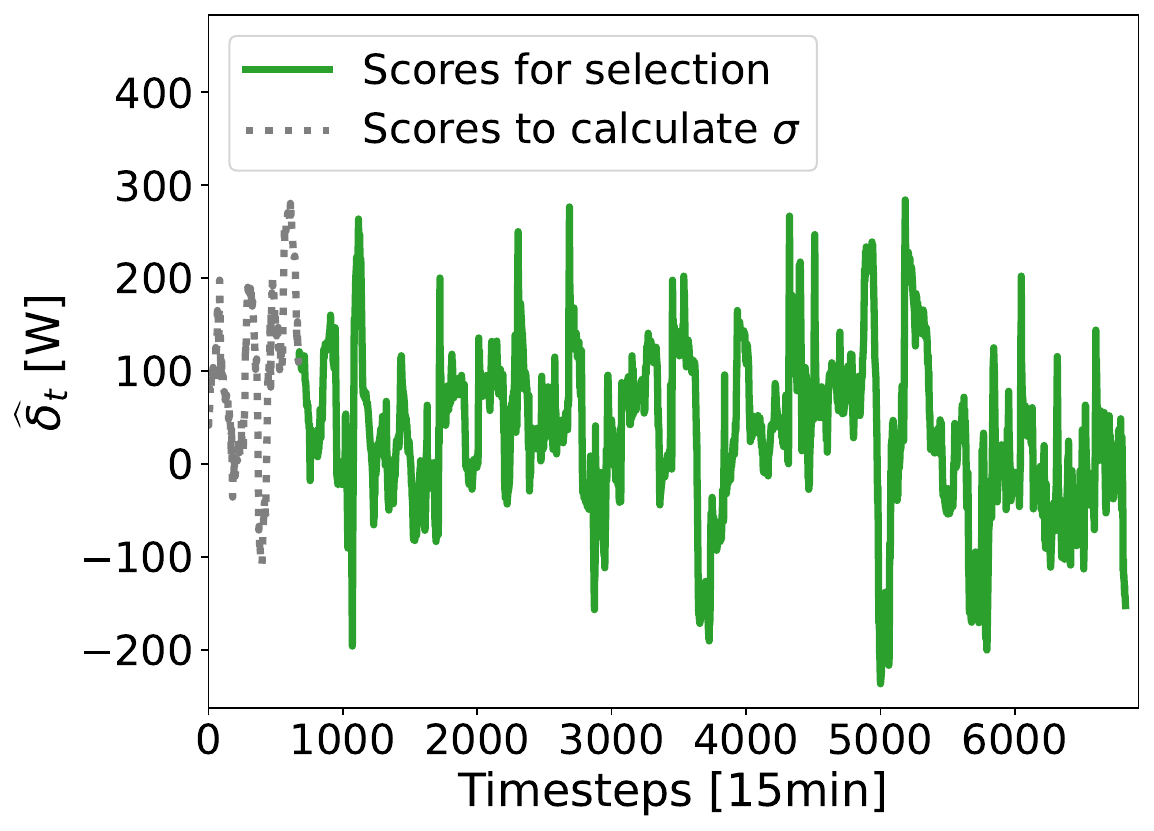}
        \label{scores_mae}
    \end{subfigure}
    \hspace{2cm}
    \begin{subfigure}[b]{0.4\textwidth}
        \centering
        \includegraphics[width=\textwidth]{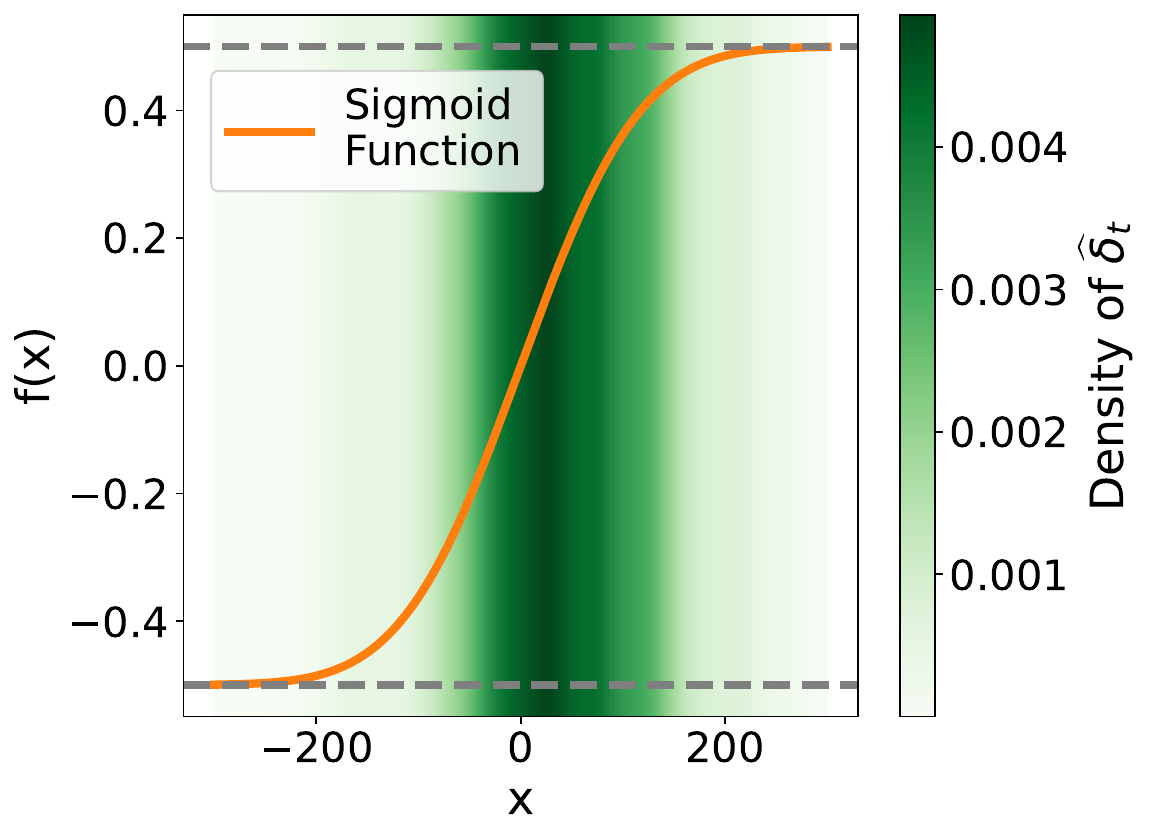}
        \label{sig_plot}
    \end{subfigure}
  \caption{\textbf{(Left)} Empirical score differences $\widehat{\delta}_t$ for the entire period. The dotted line represents the first week, which is used to calculate the scale $\sigma$. The remaining score differences are utilized for the selection procedure. \textbf{(Right)} Sigmoid function $f(x)$ in the range of the score differences $\widehat{\delta}_t$. The horizontal lines (dotted) represent the bounds at $-1/2$ and $1/2$. The color bar visualizes the quantity distribution density of the score differences $\widehat{\delta}_t$ along the x-axis.}
    \label{transformation}
    
\end{figure}

The score differences $\widehat{\delta}_t$ for the entire data set are shown in Figure~\ref{transformation} (left). 
$\sigma$ is calculated as the standard deviation of the first week of $\widehat{\delta}_t$  (dotted line in Figure~\ref{transformation} (left)). The remaining 6154 score differences are used for the selection procedure.\\

By definition, the transformed score differences $\widetilde{\delta}_t = f(\widehat{\delta}_t)$ fulfill the condition:
\begin{equation}
     \vert \widetilde{\delta}_t \vert \leq \frac{1}{2} \text{ for all timesteps t}.
\end{equation}
Figure~\ref{transformation} (right) represents the sigmoid function of Equation~\eqref{sigmoid_eq}. The dotted horizontal lines indicate the bounds at $-1/2$ and $1/2$. The color gradations visualize the quantity of score differences $\widehat{\delta}_t$ corresponding to to the histogram on the right hand side of Figure~\ref{hist_scores}. From this
we can visually verify that indeed $f$ satisfies condiitons~(i)--(iii): Most of the score differences are centered around $0$, with a slight rightward shift. This region is shaded dark green, indicating where the majority of score differences lie. Within this area, the sigmoid function is approximately linear as required. In contrast, outside this central region, represented by the lightly shaded green areas, the function exhibits strong curvature, approaching the limits of 
$\pm 1/2$ at both ends.

We aim at comparing the selection using $e$-values in Section~\ref{sec:results} with the performance of the individual models NG-RC and LSTM. Additionally, we define an ``oracle benchmark'', which always selects the method with the lower MAE. Since the oracle represents a perfect selection. 
It establishes a lower bound for the best possible score achievable by our $e$-selection procedure. The average scores of the different benchmarks during the selection period (see Figure~\ref{transformation} (left)) are presented in Table~\ref{tab:benchmark}.

\begin{table}[H]
    \centering
    \begin{tabular}{|l||c|c|c|}
      \hline 
      Method  & Oracle & NG-RC & LSTM\\
      \hline 
      Average & 425.59 W & 476.07 W & 444.38 W\\
      \hline 
    \end{tabular}
    \caption{Average scores on the whole test set of the benchmark models consisting of Oracle, NG-RC, and LSTM in Watts.}
    \label{tab:benchmark}
\end{table}

%% file: results.tex
In this section, we present the results of applying the $e$-selection procedure outlined in Section~\ref{procedure} to the electricity demand time series data described in Section~\ref{data description}. 
Specifically, we use the transformed score differentials $\widetilde{\delta}_t$ of the electricity demand forecasts and the $e$-process methodology to select a forecast model for every point in time.
Doing so, we create a combined forecast which is benchmarked against each of the individual forecasting methods and to the oracle, which represents the best possible combined prediction. 
Our analysis focuses on the persistence $e$-selection method, including a hyperparameter optimization for $\lambda$ and $\omega$. Additionally, we report the computational time required for the selection processes. \\

\begin{figure}[H]
\begin{subfigure}[t]{0.4\textwidth}
    \includegraphics[height= 0.35\textheight]{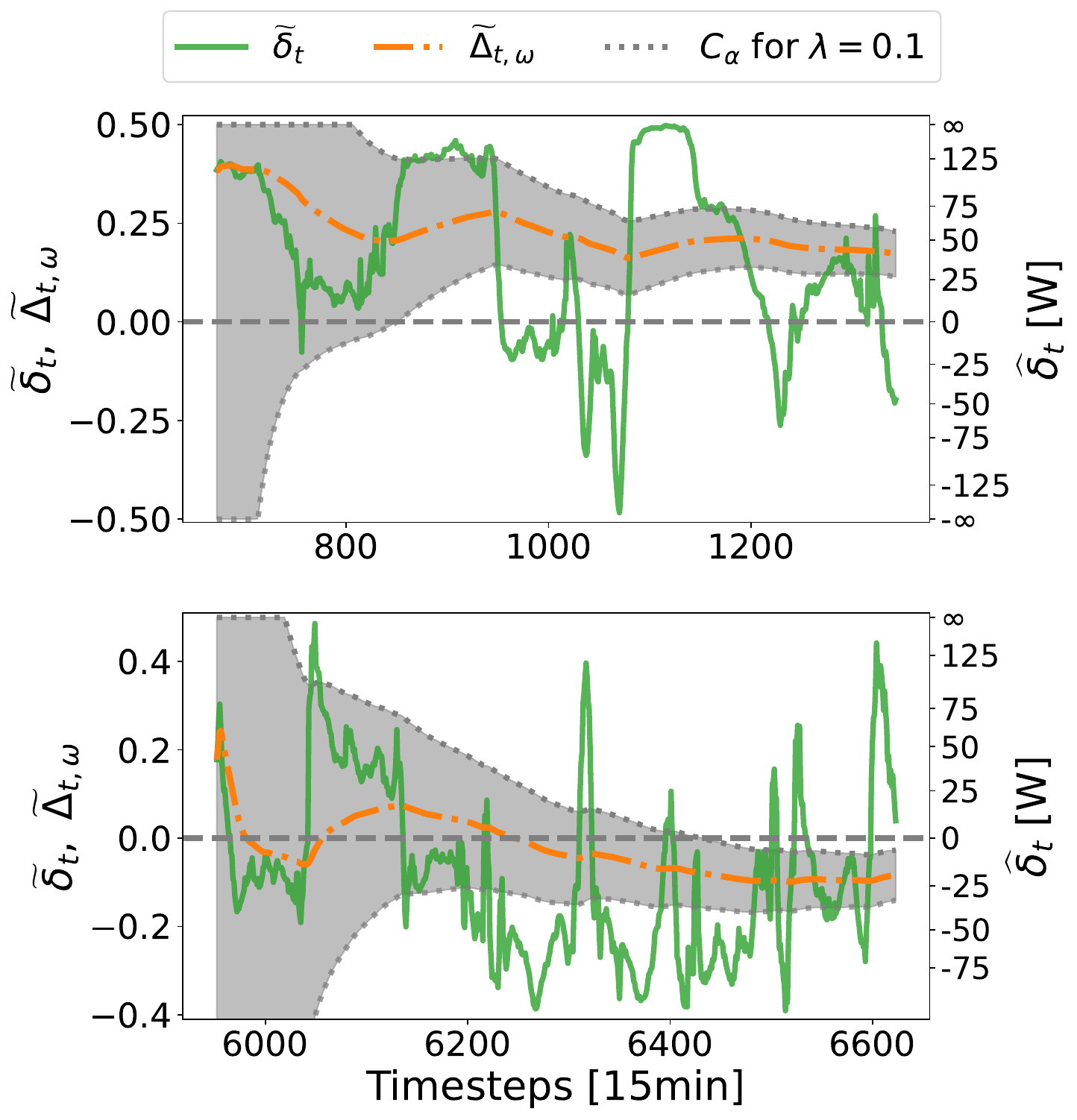}
    \centering
\label{fig:score_differentials}
\end{subfigure}
\hspace{2.5cm}
\begin{subfigure}[t]{0.4\textwidth}
    \includegraphics[height=0.35 \textheight]{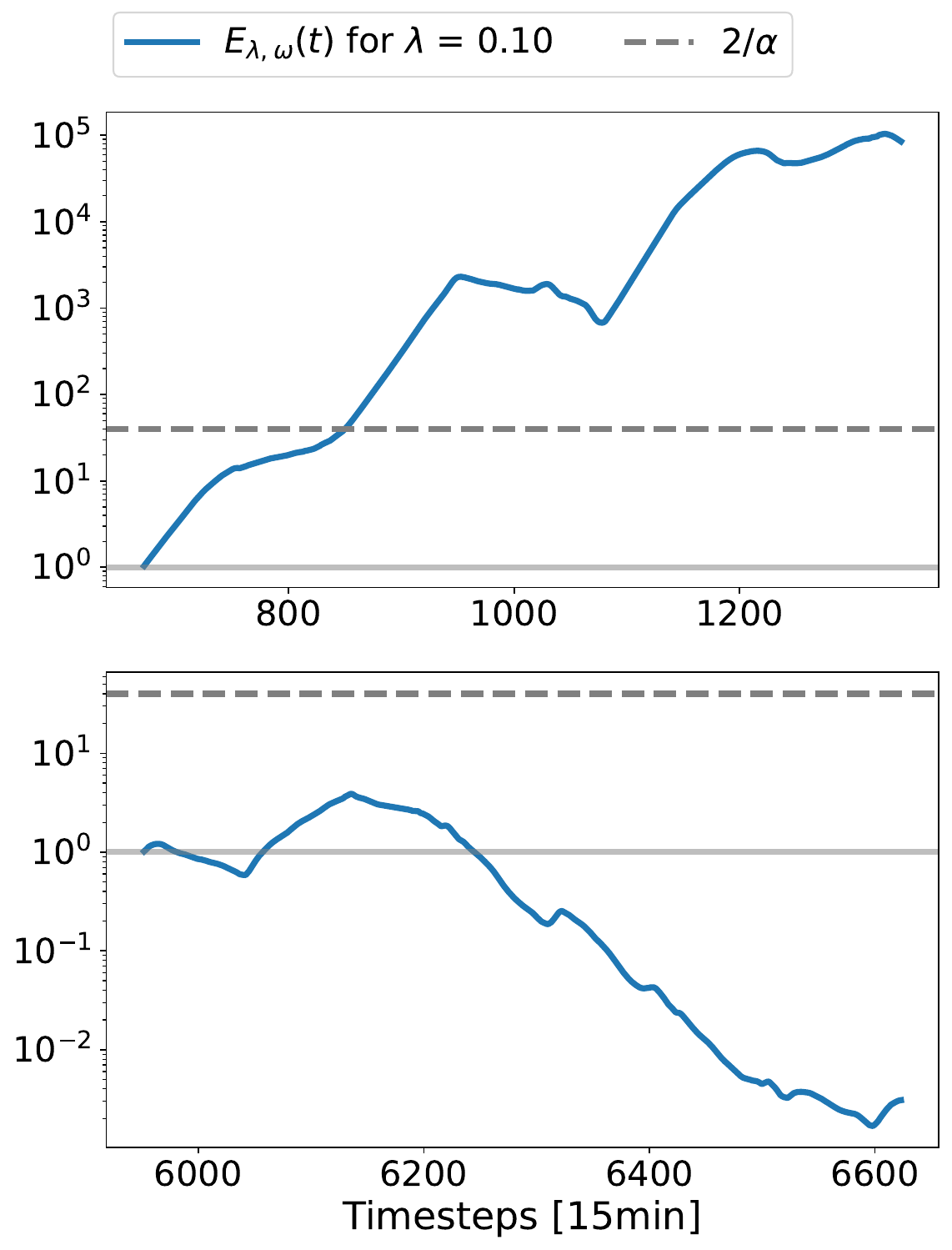}
    \label{fig:e_process}
\end{subfigure}
\caption{Comparison of different processes for two time periods:  days 7 to 14 (upper plot) and day 62 to 69 (lower plot). Confidence sequence and threshold at significance level $\alpha = 0.05$. \textbf{(Left)} Transformed score differentials $\widetilde{\delta}_t$ (solid line) and average transformed score differentials $\widetilde{\Delta}_{t, \omega}$ (dashed dotted line), along with the corresponding confidence sequence $C_\alpha$ (shaded area). The second y-axis represents the rescaled values from the first y-axis in Watts, showing a linear relationship around zero and a compression of higher deviations. The horizontal dashed line represents the value zero. \textbf{(Right)} $e$-process $E_{\lambda, \omega}(t)$ (solid line) with the threshold value $2/\alpha$ (dashed line) on a logarithmic scale. The horizontal solid line represents the starting value one. Exceeding the threshold indicates that LSTM outperforms NG-RC during this period.}
\label{fig:processes}
\end{figure}

Starting with the transformed score differentials $\widetilde{\delta}_t$, we first present exemplary results for a specific parameter combination of $\lambda$ and $\omega$. Specifically, we set $\lambda = 0.1$ and consider the data from the previous 7 days, corresponding to $\omega = 672$. 
Figure~\ref{fig:processes} (left) displays the processes $\widetilde{\delta}_t$ (solid line) and $\widetilde{\Delta}_{t, \omega}$ (dotted line), along with the corresponding confidence intervals $C_\alpha$ for $\Delta_{t, \omega}$ calculated according to Eq.~\eqref{eq:confidence_interval}, at significance level of $\alpha = 0.05$ and $\lambda = 0.1$. 
The results are shown for two time periods of the data set: The days 7 to 14 of the data set are shown in the graphs at the top of Figure~\ref{fig:processes} while the graphs at the bottom depict the days 62 to 69. 
The left y-axis in Figure~\ref{fig:processes} (left) refers to the transformed scores, while the second y-axis reverts the scale of the bounded scores to the actual scores in Watts, indicating a linear transformation around zero, with higher absolute score deviations being compressed. 
The score differentials $\widetilde{\delta}_t$ exhibit significant variability, while the average score differentials $\widetilde{\Delta}_t$ appear to converge for each time period.
The width of the confidence intervals decreases over time, eventually leading to the entire interval lying either above or below zero.\\

In Figure~\ref{fig:processes} (right), the $e$-processes $E_{\lambda, \omega}(t)$ for the same time periods are plotted with the threshold value $2/\alpha$ on a logarithmic scale. 
In the first time period, the process exceeds the threshold at time step 852, indicating the rejection of the hypothesis $\mathcal{H}_0(p,q)$. This suggests that the forecasting method LSTM outperforms NG-RC during this time period. 
The same interpretation is supported by the confidence interval, which remains entirely above zero after this time.
For the second time period, the $e$-process remains below the threshold and becomes very small. Therefore, we cannot reject the hypothesis $\mathcal{H}_0(p,q)$. To assess whether NG-RC outperforms LSTM, we should examine the process $E^*_{\lambda, \omega}(t)$ or consider the confidence interval, which remains entirely below zero after time step 6433. This indicates that NG-RC indeed outperforms LSTM during this period.\\

\begin{figure}[H]
    \centering
    \includegraphics[width=\textwidth]{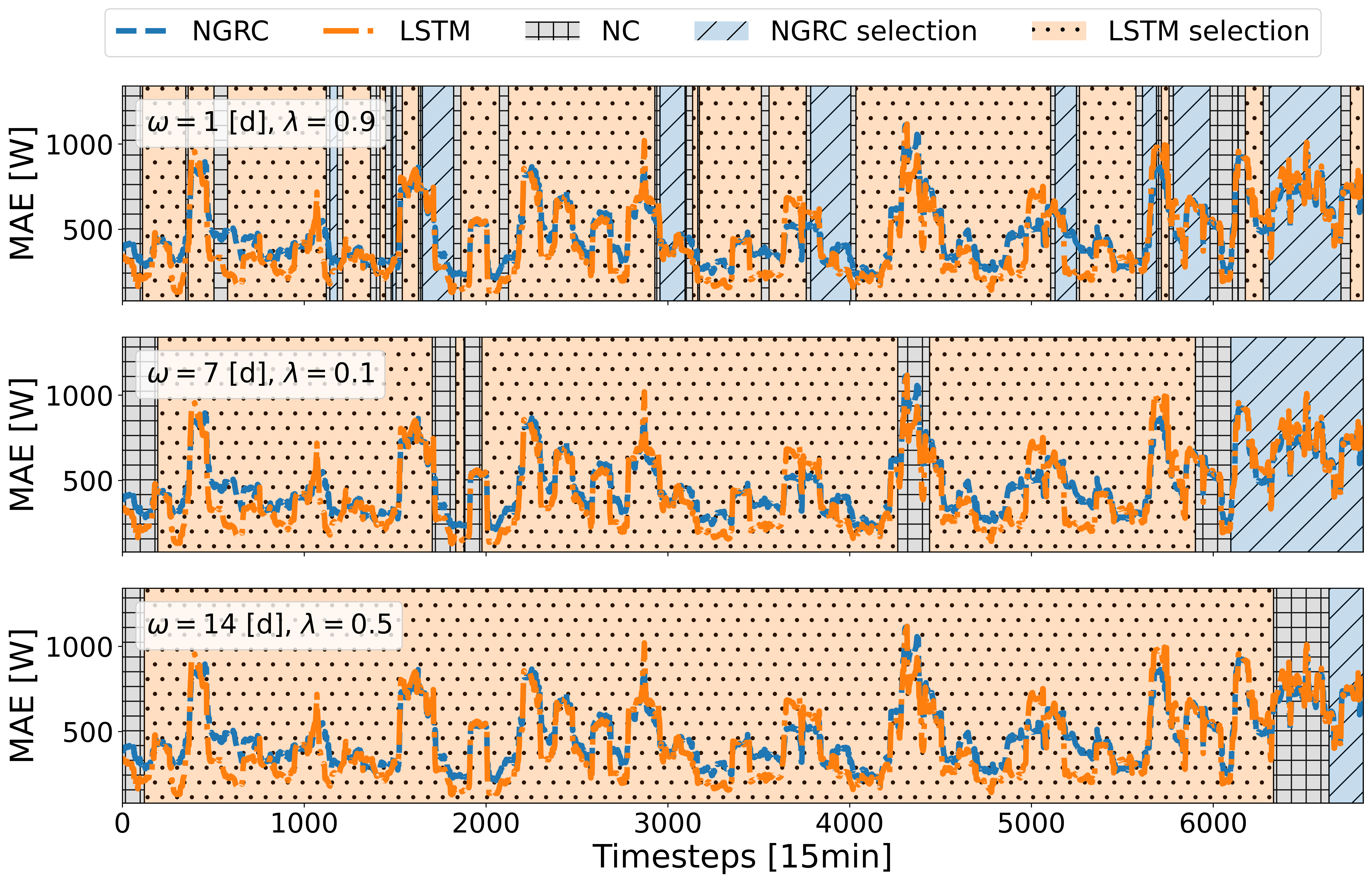}
    \caption{MAEs for NG-RC (dashed line) and LSTM (dashed dotted line) over 6154 prediction points, excluded the first week of data. Shaded areas indicate model selection by the $e$-selection method for different hyperparameters $\omega$ and $\lambda$: diagonal lines for NG-RC, dotted for LSTM. Squares show periods where no clear preference is given.}
    \label{fig:combined_selection_plots}
\end{figure}

Applying the procedure to the entire dataset of 6154 prediction points results in an $e$-selection forecast $\theta_t$ for $t \in \{1, \ldots, 6154\}$.  Figure~\ref{fig:combined_selection_plots} illustrates the MAE of the NG-RC (dotted line) and LSTM (dashed dotted line) forecasts over the entire time period and depicts three different $e$-selection processes with varying $\omega$ and $\lambda$. 
The background area style in all the three sub-figures indicates which model is selected by the $e$-process in that specific time period. 
Areas with diagonal lines represent the selection of NG-RC, while the dotted area indicates the selection of LSTM. 
Areas shaded with squares indicate time steps where no forecasting method is preferred, requiring the application of one of the approaches outlined in Section~\ref{procedure}. Note that the first week is excluded from the predictions, as it is used to calculate the scale parameter $\sigma$ for the transformation function $f(x)$ (Eq.~\eqref{sigmoid_eq}). In every plot, LSTM is the most frequently selected model. 
However, NG-RC is primarily chosen towards the end of the time series for each plot. 
Notably, smaller rolling windows $\omega$ and higher risk tolerances $\lambda$ result in more frequent and faster switches between models. Conversely, a larger rolling window of 14 days typically leads to only a single switch, occurring at the end of the time series. \\

To consider the computational cost, the runtime for each approach for the entire dataset is presented in Table~\ref{tab:runtime} across various window sizes $\omega$. All computations were performed on a Windows Server 2019 machine equipped with an Intel Xeon E5-2630v4 CPU and 256 GB DDR4-2400 RAM. The code was executed with Python 3.10, running on a single core without utilizing multiprocessing.
 Larger windows result in longer runtimes due to the increased computational demand of summing over more time steps. 

\begin{table}[H]
    \centering
    \begin{tabular}{|c||c|c|c|}
         \hline 
         \multicolumn{1}{|l||}{ } & \multicolumn{3}{|c|}{runtime $e$-selection [s]} \\ 
         \hline 
       $\omega$ [days] & Persistence & Sampling & wAvg \\
      \hline
      1  & 2.46 & 6.36 & 2.46 \\
      \hline 
      7  & 7.70 & 7.79 &  7.80\\
      \hline 
      14 & 12.50 & 12.72 & 13.12\\
      \hline 
    \end{tabular}
    \caption{Runtime of the different selection methods for various $\omega$.}
  \label{tab:runtime}
\end{table}
Due to the relatively low computational costs of applying $e$-processes for model selection, we perform a hyperparameter optimization considering $\omega$ and $\lambda$. This optimization is carried out using both a simple grid search and the python module \texttt{optuna} \cite{akiba2019optuna}. The parameter ranges considered are:
\begin{itemize}
\item $\omega \in \{1\text{h},~ 2\text{h},~1\text{d},~ 2\text{d}, \dots, 14\text{d}\}$, where h denotes the hours and d denotes the days,
\item $\lambda \in \{0.01,\dots, 0.99\}$.
\end{itemize}

For each combination, the overall average score across the entire prediction period is calculated for each of the three $e$-selection methods. Combinations where no conclusion could be reached after the first time step of the first week were excluded from consideration to maintain consistency in comparison with the persistence method.
Selected results for $\omega \in \{1\text{d},~4\text{d},~ 7\text{d},~ 14\text{d}\}$ and $\lambda \in \{0.1,~ 0.5,~ 0.9\}$ are presented in Table~\ref{results_table}. For comparison, we benchmark these scores against the average scores of NG-RC and LSTM in Table~\ref{tab:benchmark}. Scores that are lower than those of NG-RC and LSTM are highlighted in the table. The best achieved score was 441.31 W using the persistence method with $\omega = 7$ days and $\lambda = 0.07$. Although a deviation of 3.07 W from the LSTM model may seem minor, this actually represents a 16.3\% improvement compared to the LSTM's deviation from the best possible score of 425.59 W (oracle). With this parameter configuration, the $e$-selection method chooses the better forecasting model in 70.91\% of the time steps.

\begin{table}[H]
    \centering
     \begin{tabular}[t]{|c|c||l|l|l|}
     \hline 
     \multicolumn{2}{|l||}{Hyperparameters} & \multicolumn{3}{|l|}{Average score [W]} \\ 
     \hline
     $\omega$ [days]  & $\lambda$ & \textit{Persistence} & \textit{Sampling} & \textit{wAvg}\\
       \hline 
       1 & 0.1 & 449.30 & 450.90 & 451.01\\
       & 0.5 &  449.35& 448.96& 449.17\\
       & 0.9 & 449.21& 448.96& 449.19 \\ 
       \hline 
         4 & 0.1 & \textbf{441.81} & 446.66 & 446.73\\
       & 0.5 &  445.56& 446.50& 446.54\\
       & 0.9 & 445.41& 446.39& 446.54 \\ 
       \hline
       7 & 0.1 & \textbf{441.48} & \textbf{442.12} & \textbf{442.26}\\
       & 0.5 & \textbf{441.74}& \textbf{442.44}& \textbf{442.41}\\
       & 0.9 & \textbf{441.67}& \textbf{443.01}& \textbf{442.96}\\ 
       \hline
        10 & 0.1 & \textbf{442.83} & \textbf{442.44}& \textbf{442.47}\\
       & 0.5 &  \textbf{442.53}& \textbf{442.45}& \textbf{442.50}\\
       & 0.9 & \textbf{442.64}& \textbf{442.48}& \textbf{442.51} \\ 
       \hline 
       14 & 0.1 & \textbf{443.96} & \textbf{443.27} & \textbf{443.30}\\
       & 0.5 & \textbf{444.04} & \textbf{443.20} & \textbf{443.28}\\
       & 0.9 & \textbf{443.97}& \textbf{443.03}& \textbf{443.20}\\ 
      \hline
\end{tabular}
\caption{Average scores in Watts for the three $e$-selection approaches from Section~\ref{procedure} across various parameter combinations. Scores lower than those of NG-RC and LSTM in Table~\ref{tab:benchmark} are highlighted.}
\label{results_table}
\end{table}

\begin{figure}[H]
    \centering
    \includegraphics[width=\textwidth]{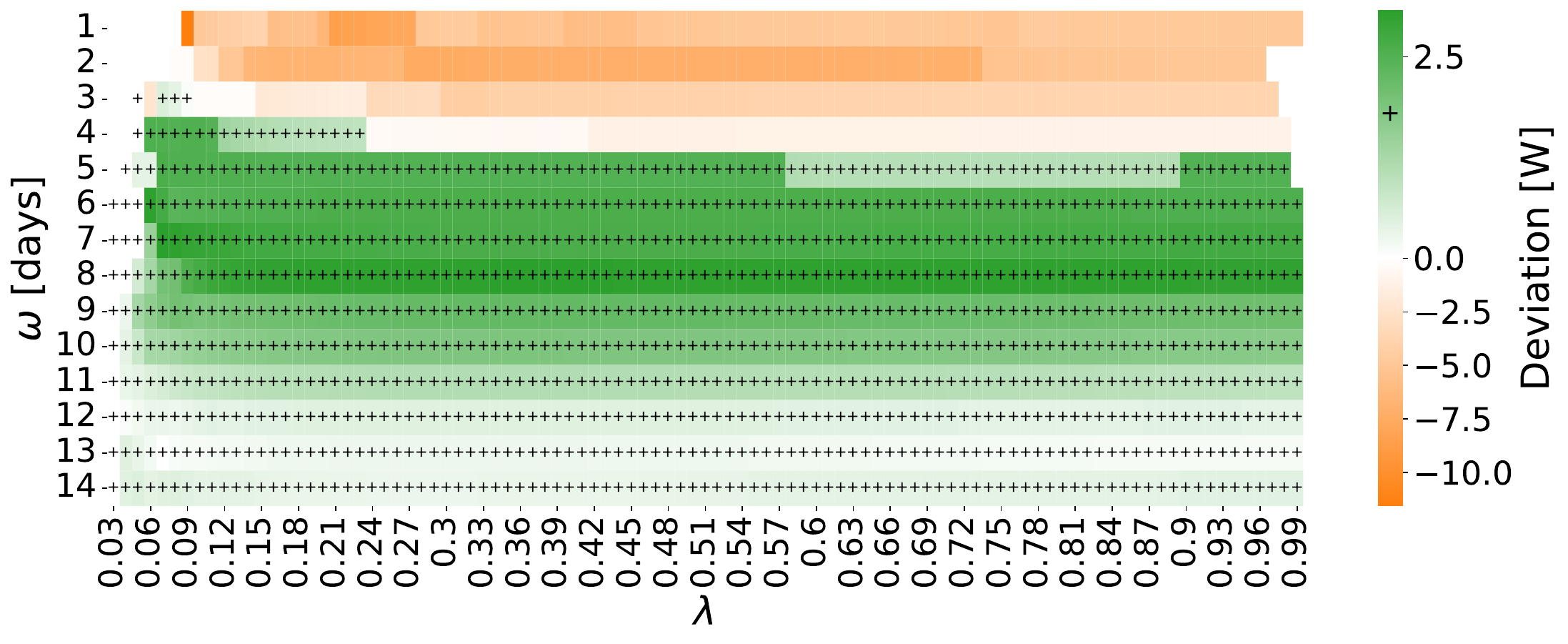}
    \caption{Deviation of LSTM and the $e$-selection persistence method of the average scores across combinations of $\omega$ and $\lambda$. Cells marked with a \glqq$+$\grqq~indicates an improvement of the performance using $e$-selection.}
    \label{fig:results_heatmap}
\end{figure}

To examine all combinations, Figure~\ref{fig:results_heatmap} shows a heatmap representing the deviation of the $e$-selection persistence method compared to the LSTM method across various parameter combinations. Combinations with window sizes of one and two hours were excluded from the heatmap because they couldn't select a method at the first time step. Cells marked with a \glqq$+$\grqq~indicate an improvement of the combined forecast  using the $e$-selection method. 
Choosing window sizes larger than five days consistently results in an improvement regardless of the choice  of $\lambda$. 
The worst performance occurred with $\lambda = 0.09$ and $\omega = 1$d, showing a worsening of the forecast performance of -11.67 W. 
Overall, in 74.24\% of the cases we get an improvement using $e$-selection persistence method. \\

For the sampling method, we observe that 64.94\% of the cases result in an improvement, with the best performance achieving a score of 441.42 W and a deviation of 2.96 W. The worst performance for this method results in a score of 450.94 W, leading to a deviation of -6.56 W. For the weighted average method, 64.87\% of the cases show an improvement, with the best deviation at 2.91 W and the worst deviation at -6.92 W. Although these methods show a slightly lower percentage of improvements compared to the persistence method, they also show significantly better worst-case performances.

%% file: discussion.tex
Our results from the previous section suggest that the $e$-selection procedure, particularly the persistence approach in our application, can outperform a fixed choice of forecasting technique. Even in the worst-case scenario, it provides an improvement over consistently choosing the NG-RC forecast. While the parameters $\lambda$ and $\omega$ control the frequency of model switching, for our data, the $e$-selection's performance appears to depend more on past $\omega$ observations than on the weights $\lambda$. Given that residential load profiles often exhibit daily and weekly patterns, it is plausible that the past $\omega = 7$ days represents the most critical time window for energy forecasting \cite{load_clustering}. 
Notably, setting $\omega = 1$ step and $\lambda \approx 1$ closely approximates a true day-ahead persistence model.
Rather than dividing the data into a separate validation set, our study demonstrates the application of $e$-values in a continuous time setting which enables real-time model selection including guarantees and known decision risk in the energy context.\\

Although computational time is not the main focus of this work, it remains crucial for real-time applications. Even though the selection procedure itself is computationally efficient, and can potentially be improved, generating two forecasts at each time step can be expensive, particularly when using computationally intensive methods like LSTM \cite{brucke2024benchmarking}.\\

Regarding the different $e$-selection approaches our example does not provide conclusive evidence favoring one method over another. For $\omega\ge 7$, where our model consistently outperforms both LSTM and NG-RC, the average scores remain extremely close,  varying by less than 4.7\% in comparison to the optimal score. 
Instead, one may focus on interpretational aspects. The persistence approach leads to longer periods using a single forecast, while the sampling and weighted average methods combine both predictions to enhance forecast accuracy.  
In this context, the weights $w_{p,t}$, can be interpreted as Bayes factors \cite{bayes_factors}. 
If a linear combination of the forecasts is allowed, the weighted average procedure is applicable. 
Otherwise, if a clear decision between $p_t$ and $q_t$
must be made at each step, the sampling approach is appropriate. \\

Despite LSTM outperforming NG-RC in the long term with respect to MAE on this dataset, NG-RC achieves a lower root mean squared error (RMSE), indicating that the selection outcome is highly dependent on the chosen scoring rule \cite{brucke2024benchmarking}. 
A key advantage of using $e$-values is their ability to control the familywise error rate in sequentially dependent data, providing guarantees for the statistical decisions. 
In our study, we test $\widetilde{\Delta}_{t, \omega}$ at each time step $t$, offering a weaker guarantee than testing the score differentials $\delta_t$ across the entire dataset. However, this guarantee applies strictly to the statistical test, not to the overall prediction, which requires additional assumptions.  
Consequently, it is challenging to verify whether the confidence level $1-\alpha$ is actually achieved \cite{choe2023comparing}.

%% file: conclusion.tex
In this work, we successfully transfer and translate the $e$-value based approach for time-continuous forecast model selection from
\cite{valid_sequential_inference_on_probability_forecast_performance,choe2023comparing} from binary outcome prediction to the energy domain.   
We extend this forecast model selection approach to forecast fusion / combination by specifying a combination of the two forecasts into a new better one. Our study demonstrates that the $e$-selection method provides competitive results with minimal additional computational costs while offering a statistical guarantee. However, it's important to note that this guarantee does not imply that the sequential test will make the correct decision at every time step with an error rate of $\alpha$, because this would involve information on the outcome distribution unknown
to the decision process. 
Instead, it controls the average score differentials with a significance level of $1-\alpha$ \cite{choe2023comparing}.
The passage from the binary outcome setting of \cite{valid_sequential_inference_on_probability_forecast_performance,choe2023comparing} to the continuous outcome setting of energy demand also requires usage of scores adapted to this scale. 
To this end, we replace the Brier scores used in \cite{valid_sequential_inference_on_probability_forecast_performance,choe2023comparing}, by the MAE which is well established in the energy context. 
To achieve this, we suitably transform the MAE scores in an order-preserving manner. For interpretability, in our decision plots, we back-transform the results and confidence bounds to the original MAE scale. 
With these two adaptions -- extension to outcomes with values on a continuous scale and usage of
a corresponding score -- we obtain SAVI tests for superiority and time-uniform confidence sequences.
On the real data example of time series of residential electricity demand, these tests are shown to have enough power to obtain a clear ordering of the considered forecasters most of the time which in addition is supported by a probabilistic guarantee, here in form of a warranted familywise significance level.\\

As indicated, the probabilistic guarantees do not necessarily extend to a retrospective backtesting perspective, so future work should focus on properly backtesting the results presented in this study. The model-free and $e$-process-based backtesting procedure introduced in \cite{wang2023ebacktesting,wang2023ebacktestingSSRN} offers a promising direction for such a validation. 
Additionally, further studies are needed to evaluate the performance and robustness of our $e$-selection method across different datasets and scoring rules. This includes validation of the hyperparameters involved. Moreover, the selection procedure should account for the varying computational costs associated with different forecasting methods. 
Passing from point forecasts to probabilistic forecasts, it is clear that one could easliy proceed the same way as in this work, simply replacing the MAE score function by the terms of a continuous ranking probability score (CRPS) as discussed in \cite{gneiting_probabilistic_forecasting}.  
Finally, instead of comparing just two forecasting techniques, future research could explore a wider range of methods, allowing for more adaptable selection procedure that handle diverse forecasting approaches.

%% file: main.bbl
\begin{thebibliography}{10}

\bibitem{amasyali2018}
K.~Amasyali and N.~M. El-Gohary, ``A review of data-driven building energy consumption prediction studies,'' {\em Renewable and Sustainable Energy Reviews}, vol.~81, pp.~1192--1205, 2018.

\bibitem{deb2017}
C.~Deb, F.~Zhang, J.~Yang, S.~E. Lee, and K.~W. Shah, ``A review on time series forecasting techniques for building energy consumption,'' {\em Renewable and Sustainable Energy Reviews}, vol.~74, pp.~902--924, 2017.

\bibitem{golmohamadi2022}
H.~Golmohamadi, ``Demand-side flexibility in power systems: A survey of residential, industrial, commercial, and agricultural sectors,'' {\em Sustainability}, vol.~14, no.~13, p.~7916, 2022.

\bibitem{babatunde2020}
O.~M. Babatunde, J.~L. Munda, and Y.~Hamam, ``Power system flexibility: A review,'' {\em Energy Reports}, vol.~6, pp.~101--106, 2020.

\bibitem{gazafroudi2018}
A.~S. Gazafroudi, F.~Prieto-Castrillo, T.~Pinto, and J.~M. Corchado, ``Energy flexibility management in power distribution systems: decentralized approach,'' in {\em 2018 International Conference on Smart Energy Systems and Technologies (SEST)}, (Seville, Spain), pp.~1--6, IEEE, 2018.

\bibitem{LPG}
N.~Pflugradt, P.~Stenzel, L.~Kotzur, and D.~Stolten, ``{LoadProfileGenerator: An Agent-Based Behavior Simulation for Generating Residential Load Profiles},'' {\em Journal of Open Source Software}, vol.~7, p.~3574, 03 2022.

\bibitem{de2009}
L.~C.~M. de~Andrade and I.~N. da~Silva, ``{Very short-term load forecasting based on ARIMA model and intelligent systems},'' in {\em 2009 15th International Conference on Intelligent System Applications to Power Systems}, (Curitiba, Brazil), pp.~1--6, IEEE, 2009.

\bibitem{guo2006}
Y.-C. Guo, D.-X. Niu, and Y.-X. Chen, ``Support vector machine model in electricity load forecasting,'' in {\em 2006 International Conference on Machine Learning and Cybernetics}, (Dalian, China), pp.~2892--2896, IEEE, 2006.

\bibitem{kong2017}
W.~Kong, Z.~Y. Dong, Y.~Jia, D.~J. Hill, Y.~Xu, and Y.~Zhang, ``{Short-term residential load forecasting based on LSTM recurrent neural network},'' {\em IEEE Transactions on Smart Grid}, vol.~10, no.~1, pp.~841--851, 2017.

\bibitem{brucke2021}
K.~Brucke, S.~Arens, J.-S. Telle, T.~Steens, B.~Hanke, K.~von Maydell, and C.~Agert, ``A non-intrusive load monitoring approach for very short-term power predictions in commercial buildings,'' {\em Applied Energy}, vol.~292, p.~116860, 2021.

\bibitem{brucke2024benchmarking}
K.~Brucke, S.~Schmitz, D.~K{\"o}glmayr, S.~Baur, C.~R{\"a}th, E.~Ansari, and P.~Klement, ``Benchmarking reservoir computing for residential energy demand forecasting,'' {\em Energy and Buildings}, p.~114236, 2024.

\bibitem{simani2024}
K.~N. Simani, Y.~O. Genga, and Y.-C.~J. Yen, ``Using lstm to perform load predictions for grid-interactive buildings,'' {\em SAIEE Africa Research Journal}, vol.~115, no.~2, pp.~42--47, 2024.

\bibitem{sun2020}
Y.~Sun, F.~Haghighat, and B.~C. Fung, ``A review of the-state-of-the-art in data-driven approaches for building energy prediction,'' {\em Energy and Buildings}, vol.~221, p.~110022, 2020.

\bibitem{guo2018}
Y.~Guo, J.~Wang, H.~Chen, G.~Li, J.~Liu, C.~Xu, R.~Huang, and Y.~Huang, ``Machine learning-based thermal response time ahead energy demand prediction for building heating systems,'' {\em Applied Energy}, vol.~221, pp.~16--27, 2018.

\bibitem{kwak2015}
Y.~Kwak, J.-H. Huh, and C.~Jang, ``Development of a model predictive control framework through real-time building energy management system data,'' {\em Applied Energy}, vol.~155, pp.~1--13, 2015.

\bibitem{guo2021}
N.~Guo, X.~Zhang, Y.~Zou, L.~Guo, and G.~Du, ``Real-time predictive energy management of plug-in hybrid electric vehicles for coordination of fuel economy and battery degradation,'' {\em Energy}, vol.~214, p.~119070, 2021.

\bibitem{quan2021}
S.~Quan, Y.-X. Wang, X.~Xiao, H.~He, and F.~Sun, ``Real-time energy management for fuel cell electric vehicle using speed prediction-based model predictive control considering performance degradation,'' {\em Applied Energy}, vol.~304, p.~117845, 2021.

\bibitem{practical_guide_model_selection}
A.~T. Tredennick, G.~Hooker, S.~P. Ellner, and P.~B. Adler, ``A practical guide to selecting models for exploration, inference, and prediction in ecology,'' {\em Ecology}, vol.~102, no.~6, p.~e03336, 2021.

\bibitem{swanson1997}
N.~R. Swanson and H.~White, ``A model selection approach to real-time macroeconomic forecasting using linear models and artificial neural networks,'' {\em Review of Economics and Statistics}, vol.~79, no.~4, pp.~540--550, 1997.

\bibitem{billah2006}
B.~Billah, M.~L. King, R.~D. Snyder, and A.~B. Koehler, ``Exponential smoothing model selection for forecasting,'' {\em International Journal of Forecasting}, vol.~22, no.~2, pp.~239--247, 2006.

\bibitem{liu2022}
J.~Liu, Z.~Yu, H.~Zuo, R.~Fu, and X.~Feng, ``Multi-stage residual life prediction of aero-engine based on real-time clustering and combined prediction model,'' {\em Reliability Engineering \& System Safety}, vol.~225, p.~108624, 2022.

\bibitem{douk}
P.~Bertail, P.~Doukhan, and P.~Soulier, {\em Dependence in Probability and statistics}.
\newblock Springer, 2006.

\bibitem{wald}
A.~Wald, ``Sequential tests of statistical hypotheses,'' {\em The Annals of Mathematical Statistics}, vol.~16, no.~2, pp.~117--186, 1945.

\bibitem{Cusum}
J.-L. Vivancos, R.~A. Buswell, P.~Cosar-Jorda, and C.~Aparicio-Fernández, ``The application of quality control charts for identifying changes in time-series home energy data,'' {\em Energy and Buildings}, vol.~215, p.~109841, 2020.

\bibitem{gametheoretic}
A.~Ramdas, P.~Gr{\"u}nwald, V.~Vovk, and G.~Shafer, ``Game-theoretic statistics and safe anytime-valid inference,'' {\em Statistical Science}, vol.~38, no.~4, pp.~576--601, 2023.

\bibitem{TestMartingalesBayesFactors}
G.~Shafer, A.~Shen, N.~Vereshchagin, and V.~Vovk, ``{Test Martingales, Bayes Factors and p-Values},'' {\em Statistical Science}, vol.~26, no.~1, pp.~84 -- 101, 2011.

\bibitem{valid_sequential_inference_on_probability_forecast_performance}
A.~Henzi and J.~F. Ziegel, ``Valid sequential inference on probability forecast performance,'' {\em Biometrika}, vol.~109, no.~3, pp.~647--663, 2022.

\bibitem{choe2023comparing}
Y.~J. Choe and A.~Ramdas, ``Comparing sequential forecasters,'' {\em Operations Research}, vol.~72, no.~4, pp.~1368--1387, 2024.

\bibitem{mae_load_forecasts}
J.~Coignard, M.~Janvier, V.~Debusschere, G.~Moreau, S.~Chollet, and R.~Caire, ``Evaluating forecasting methods in the context of local energy communities,'' {\em International Journal of Electrical Power \& Energy Systems}, vol.~131, p.~106956, 2021.

\bibitem{SafeTesting}
P.~Gr{\"u}nwald, R.~de~Heide, and W.~Koolen, ``Safe testing. {D}iscussion paper.,'' {\em Journal Journal of the Royal Statistical Society. Series B: Statistical Methodology}, Accepted/In press - 24 Jan 2024.
\newblock Accepted author version available under \url{https://research.utwente.nl/files/353763913/Preprint-Grunwald-24-Jan-2024.pdf}.

\bibitem{ramdas2022admissible}
A.~Ramdas, J.~Ruf, M.~Larsson, and W.~Koolen, ``Admissible anytime-valid sequential inference must rely on nonnegative martingales,'' 2022.
\newblock Available on ArXiV under \url{https://arxiv.org/abs/2009.03167}.

\bibitem{probability_theory}
A.~Klenke, {\em Probability Theory : A Comprehensive Course}.
\newblock Springer International Publishing, 3rd~ed., 2020.

\bibitem{betting_score}
G.~Shafer, ``{Testing by Betting: A Strategy for Statistical and Scientific Communication},'' {\em Journal of the Royal Statistical Society Series A: Statistics in Society}, vol.~184, pp.~407--431, 05 2021.

\bibitem{vovk2021}
V.~Vovk and R.~Wang, ``E-values: Calibration, combination and applications,'' {\em The Annals of Statistics}, vol.~49, no.~3, pp.~1736--1754, 2021.

\bibitem{MultipleTesting}
R.~Miller, {\em Simultaneous Statistical Inference}.
\newblock Springer, 2nd~ed., 1981.

\bibitem{HowardEtAlTimeUnifConf}
S.~R. Howard, A.~Ramdas, J.~McAuliffe, and J.~Sekhon, ``Time-uniform, nonparametric, nonasymptotic confidence sequences.,'' {\em The Annals of Statistics}, vol.~49, no.~2, p.~1055–1080, 2021.

\bibitem{howard2020timeuniform}
S.~R. Howard, A.~Ramdas, J.~McAuliffe, and J.~Sekhon, ``Time-uniform chernoff bounds via nonnegative supermartingales,'' {\em Probability Surveys}, no.~17, pp.~257--317, 2020.

\bibitem{multiple_testing}
J.~P. Romano, A.~M. Shaikh, M.~Wolf, {\em et~al.}, ``Multiple testing,'' {\em The new Palgrave dictionary of economics}, vol.~4, 2010.

\bibitem{strictly_proper}
T.~Gneiting and A.~E. Raftery, ``Strictly proper scoring rules, prediction, and estimation,'' {\em Journal of the American Statistical Association}, vol.~102, no.~477, pp.~359--378, 2007.

\bibitem{fusion}
C.~Gilbert, J.~Browell, and B.~Stephen, ``Probabilistic load forecasting for the low voltage network: Forecast fusion and daily peaks,'' {\em Sustainable Energy, Grids and Networks}, vol.~34, p.~100998, 2023.

\bibitem{dataset}
D.~Musikhina, J.~Seidemann, and S.~Feilmeier, 2021.
\newblock EMSIG: Energy Management System, Data, available at:, \url{https://openenergyplatform.org/dataedit/view/demand/emsig_energy_data_by_ems} (2024).

\bibitem{ngrc_paper}
D.~J. Gauthier, E.~Bollt, A.~Griffith, and W.~A.~S. Barbosa, ``Next generation reservoir computing,'' {\em Nature Communications}, vol.~12, no.~1, p.~5564, 2021.

\bibitem{ng_rc_paper2}
I.~Ratas and K.~Pyragas, ``Application of next-generation reservoir computing for predicting chaotic systems from partial observations,'' {\em Phys. Rev. E}, vol.~109, p.~064215, Jun 2024.

\bibitem{lstm}
S.~Hochreiter and J.~Schmidhuber, ``{Long Short-Term Memory},'' {\em Neural Computation}, vol.~9, pp.~1735--1780, 11 1997.

\bibitem{load_forecast_review}
T.~Hong and S.~Fan, ``Probabilistic electric load forecasting: A tutorial review,'' {\em International Journal of Forecasting}, vol.~32, no.~3, pp.~914--938, 2016.

\bibitem{akiba2019optuna}
T.~Akiba, S.~Sano, T.~Yanase, T.~Ohta, and M.~Koyama, ``{O}ptuna: A next-generation hyperparameter optimization framework,'' in {\em The 25th ACM SIGKDD International Conference on Knowledge Discovery \& Data Mining}, pp.~2623--2631, 2019.

\bibitem{load_clustering}
S.~Ryu, H.~Choi, H.~Lee, H.~Kim, and V.~W.~S. Wong, ``Residential load profile clustering via deep convolutional autoencoder,'' in {\em 2018 IEEE International Conference on Communications, Control, and Computing Technologies for Smart Grids (SmartGridComm)}, pp.~1--6, 2018.

\bibitem{bayes_factors}
L.~Held and M.~Ott, ``On p-values and bayes factors,'' {\em Annual Review of Statistics and Its Application}, vol.~5, no.~Volume 5, 2018, pp.~393--419, 2018.

\bibitem{wang2023ebacktesting}
Q.~Wang, R.~Wang, and J.~Ziegel, ``E-backtesting,'' 2024.
\newblock Available on ArXiV under \url{https://arxiv.org/abs/2209.00991}.

\bibitem{wang2023ebacktestingSSRN}
Q.~Wang, R.~Wang, and J.~Ziegel, ``Simulation and data analysis for e-backtesting,'' 2023.
\newblock Available at SSRN under \url{https://ssrn.com/abstract=4346325}.

\bibitem{gneiting_probabilistic_forecasting}
T.~Gneiting and M.~Katzfuss, ``Probabilistic forecasting,'' {\em Annual Review of Statistics and Its Application}, vol.~1, no.~1, pp.~125--151, 2014.

\end{thebibliography}
